\documentclass[preprint,3p,12pt,times,nonatbib]{elsarticle}

\usepackage{tabularx}
\usepackage{booktabs}

\usepackage[T1]{fontenc}
\usepackage[utf8]{inputenc}

\usepackage{newtxtext,newtxmath}

\usepackage{amssymb}

\usepackage{amsmath}
\usepackage{bm}

\usepackage{graphicx}
\usepackage{epstopdf}
\usepackage{geometry}
\usepackage{placeins}
\usepackage{float}

\usepackage{subcaption}
\usepackage{caption}

\usepackage{setspace}
\usepackage{comment}
\usepackage{xcolor}
\usepackage{soul}

\usepackage{algorithm}
\usepackage{algpseudocode}

\usepackage[numbers]{natbib}

\usepackage[
colorlinks=true,
citecolor=blue,
linkcolor=black,
urlcolor=cyan
]{hyperref}

\usepackage[capitalise]{cleveref}

\graphicspath{{Figures/}}

\newcommand{\sch}[1]{\textcolor{magenta}{(#1)}}

\journal{Elsevier} 

\begin{document}

\begin{frontmatter} 

\title{Studying Creep-Fatigue interaction of Nickel-Based Superalloys using Crystal Plasticity and Entropy-Based life prediction model}


\author[aff1]{Santosh Kumar Shaw}
\ead{amz228601@am.iitd.ac.in}

\author[aff1]{Sabyasachi Chatterjee\corref{cor1}}
\ead{sabyasachi@am.iitd.ac.in}

\author[aff2]{Ayan Bhowmik}
\ead{Ayan.Bhowmik@mse.iitd.ac.in}

\author[aff3]{Alankar Alankar}
\ead{alankar@iitb.ac.in}

\cortext[cor1]{Corresponding authors}

\address[aff1]{Department of Applied Mechanics, Indian Institute of Technology Delhi, New Delhi 110016, India}
\address[aff2]{Department of Materials Science and Engineering, Indian Institute of Technology Delhi, New Delhi 110016, India}
\address[aff3]{Department of Mechanical Engineering, Indian Institute of Technology Bombay, Mumbai 400076, India}
\begin{abstract}


 Creep-fatigue interaction in single-crystal nickel superalloys is difficult to predict because the response depends on the combined effects of loading parameters, hold time, temperature, and the underlying deformation mechanisms. This is important for turbine blade applications, where components experience both fatigue and creep during service. In the present work, a crystal plasticity finite element (CPFE) framework is used to study the creep-fatigue response of a single-crystal nickel superalloy under a range of practically relevant thermo-mechanical loading conditions. In particular, the effects of strain amplitude, R-ratio, hold duration, and temperature on cyclic deformation, stress relaxation, damage evolution, and creep-fatigue life are examined. Particular attention is given to separate the roles of fatigue and creep damage, understanding their interaction, and identify the creep-dominated and fatigue-dominated regimes as a function of strain amplitude and hold time. The study brings together these effects within a single framework and shows that the predicted trends in cyclic response and life are in good agreement with experimental observations reported in the literature.

\end{abstract}


\end{frontmatter} 



\section{\textbf{Introduction}}


Nickel superalloys are a class of advanced material suitable for high temperature application such as turbine blades in gas turbines and jet engines. This is primarily due to the presence of $\gamma'$ phase of $Ni_3 Al$ which acts as obstacles to dislocation motion, significantly increasing the strength even at high temperatures. Single Crystal Nickel superalloys are particularly favourable due to absence of grain boundaries, which reduces creep damage. Although much effort has been made to understand and model the response of nickel superalloys to individual creep and cyclic loading, relatively limited studies have been conducted to understand the interaction of creep and fatigue. This is important since turbine blades are seldom subjected to pure fatigue or creep but usually experience a combination of both during a loading cycle. A typical loading cycle consists of 3 stages: (1) ramp from 0 to maximum rotation speed (2) operating under the maximum rotation speed for a duration called dwell or hold time (3) ramp from maximum rotation speed to 0. Rotation causes centrifugal loads and hence, the ramp loading/unloading phase corresponds to fatigue loading while the operation under constant rotation corresponds to creep.  Modelling the mechanical response and prediction of life for such critical components therefore requires careful consideration of the creep-fatigue interaction during loading cycles. 

Creep-fatigue interaction in Nickel superalloys and its effect on service life has been studied using both experimental and modelling approaches. Strain controlled Low cycle fatigue (LCF) tests and Creep-Fatigue Interaction tests were conducted on polycrystalline nickel superalloy samples at 650$^\circ$C by \cite{CHEN2016175}. A significant reduction in life with  increase in strain amplitude in pure fatigue and with increase in tensile hold time in Creep-Fatigue tests was reported by the authors. \cite{Shi2013} conducted extensive experiments to study the effect of orientation, temperature and dwell time on the LCF life of single crystal nickel superalloy samples. Such experimental studies involve fitting the LCF and CFI life data to empirical models and hence are limited in their predictive ability. Firstly, they do not account for the physics that govern the complex microstructural deformation mechanisms of such materials, which significantly affects the fatigue crack initiation time. Secondly, components such as turbine blades have complex geometry, which are very different from lab samples. Hence, stress concentration develops at locations such as root and trailing edges which are frequent sites of crack nucleation. Empirical models cannot account for this spatial variation of stress and strain in the material. 

These limitations have led to the development of crystal plasticity finite element (CPFE) based approaches. Although extensive studies have been conducted using CPFE for pure fatigue, its application to study creep-fatigue interaction has been relatively more recent. \cite{ESTRADARODAS201814} and \cite{Ranjan2021}  proposed detailed crystal plasticity models which considers separate deformation mechanisms for the $\gamma$ and $\gamma'$ phases. These models account for temperature, rate and orientation dependence in creep-fatigue behaviour across wide range of temperature, while also incorporating microstructural information like volume fraction and size of precipitates. However, owing to the complex deformation mechanisms and dislocation behaviour in Nickel superalloys, some differences exist between experimental data and predictions in these work. These models are also cumbersome, involving large number of evolution equations and parameters, making such models both computationally expensive and difficult to calibrate across wide range of operating conditions. \cite{STAROSELSKY20112060} proposed a novel rate-independent plasticity model in which the plastic shear rate is composed of time dependent creep and rate independent plastic components. However, the model is not capable of predicting strain rate dependence and is yet to be tested for larger strain amplitudes. 

It has long been established that accumulated slip and stored elastic energy are important indicator of fatigue crack nucleation in \cite{WAN201490} and \cite{CHEN2018213}. Although the success of such parameters are widely reported in the case of fatigue life predictions, relatively limited damage indicator parameters for creep-fatigue life have been proposed yet. Many researchers (\cite{cao_zhang_IJP_unified_2021}, \cite{zhang2017interaction}) proposed that adding a static recovery term to the kinematic hardening of Chaboche type model is able to capture the effect of stress relaxation and during hold times in strain controlled CFI tests and ratcheting during stress controlled CFI tests. Building on these observations, \cite{WANG2020105879} proposed a simplified and lightweight phenomenological CPFE model to predict creep-fatigue crack initiation life in polycrystalline Nickel superalloy. They proposed individual fatigue and creep damage indicator parameters based on accumulated slip and accumulated plastic dissipation and showed improved agreement between experimental and predicted creep fatigue life using Linear Damage Summation (LDS) and Nonlinear Damage Summation (NDS) rules, compared to traditional Coffin-Manson type models. A similar model proposed by \cite{lu2023crystal} for single crystal Nickel superalloy incorporates further microstructural details such as precipitate size and volume fraction in the slip system strength and dislocation density based parameters in the kinematic hardening rule. They also utilized an entropy based life prediction model based on the work of \cite{NADERI2010875}, demonstrating its predictive capability. However, their work is limited to limited cases such as fully reversed loading at constant temperature and requires further experimental validation.


The objective of the present work is a systematic CPFE study of the creep-fatigue response of a single-crystal nickel superalloy under a range of practically relevant thermo-mechanical loading conditions. The effects of strain amplitude, R-ratio, hold duration, and temperature on cyclic deformation, stress relaxation, damage evolution, and creep-fatigue life are examined in a unified manner, within a single framework. Special attention is given to separating the respective roles of fatigue and creep damage, understanding their interaction, and identifying the dominant damage regime as a function of strain amplitude and hold time.

 The paper is organized as follows. The CPFE framework and the life prediction model utilized in this work are described in Section 2. The model implementation and calibration details are described in Section 3. The results using the model are reported and discussed in Section 4. Finally, the conclusion of the work is presented in Section 5.  

\section{Methodology}
\label{sec:method}

\subsection{Crystal plasticity framework}\label{sec:cp}
The constitutive modelling framework is based on the dislocation density-based microstructure-sensitive  work
of Lu et al. (2023), where a crystal plasticity model was proposed
to model the creep, fatigue and  creep-fatigue interaction behaviour of nickel-based superalloy.

This framework is based on the multiplicative decomposition of the deformation gradient $\mathbf{F}$ into the elastic deformation gradient $\mathbf{F}^e$ and the plastic deformation gradient $\mathbf{F}^p$ as \cite{lee1969FeFp}:
\begin{equation}
    \mathbf{F} = \mathbf{F}^e \cdot \mathbf{F}^p
\end{equation}
where the intermediate configuration is obtained by transforming the reference configuration through the plastic deformation gradient $\mathbf{F}^p$ . Then the intermediate configuration was transformed into the current configuration through elastic deformation and rigid body rotation $\mathbf{F}^e$. The plastic velocity gradient $\mathbf{L}_p$ is defined in the intermediate configuration by summing the shear strain rates $\dot\gamma^{\alpha}$ of all available slip systems:

\begin{equation}
     \mathbf{L}^p = \dot{\mathbf{F}}^{{p}^{}} \mathbf{F}^{{p}^{-1}} = \sum_{\alpha=1}^{N} \dot{\gamma}^\alpha (\boldsymbol{s}_0^\alpha \otimes \boldsymbol{m}_0^\alpha)
\end{equation}
where $\mathbf{s}^\alpha$ and $\mathbf{m}^\alpha$ denote the unit vectors representing the slip direction and the slip plane normal respectively, in the reference configuration.

The relationship between the shear strain rate $\dot\gamma^{\alpha}$ and the resolved shear stress $\tau^\alpha$ on a given slip system is governed by power-law flow-rule proposed by \cite{Hutchinson1976}:
\begin{equation}\label{eq:sliprate}
    \dot{\gamma}^\alpha = \dot{\gamma}_0 \left| \frac{\tau^\alpha - \chi^\alpha}{g^\alpha} \right|^{\frac{1}{m}} \text{sgn}(\tau^\alpha) . 
\end{equation}
In this expression, $\dot{\gamma}^\alpha$ and $\tau^\alpha$ denote the shear strain rate and the resolved shear stress on slip system $\alpha$, respectively. The internal variables $g^\alpha$ and $\chi^\alpha$ represent the slip resistance and the back stress, while $m$ corresponds to the strain rate sensitivity exponent.

The slip resistance $g^\alpha$ is composed of initial strength $g_0^\alpha(T)$ and a component which accounts for strengthening due to dislocation hardening $g_{for}^\alpha$: 
\begin{equation}
   g^\alpha=g_0^\alpha(T) +g_{for}^\alpha(T). 
\end{equation}

The initial slip resistance $g_0^\alpha$, is dependent on temperature and strain rate using the following expression \cite{kocks1975thermodynamics,HullBacon2011} :

\begin{equation}
    \frac{g_0^\alpha(T)}{g_0^\alpha(0)} = \frac{k T}{\Delta F} \ln \left( \frac{\dot{\epsilon}}{\dot{\epsilon}_{0,t}} \right) + 1
\end{equation}

where $k/\Delta F$ is treated as a material constant, $T$ is the homologous temperature, $\dot{\epsilon}$ is the applied strain rate, $\dot{\epsilon}_{0,t}$ is the reference strain rate at 0 K, and $g_t^\alpha(0)$ is the stress required to overcome the obstacle at 0 K, whose  value is provided in the \cref{tab:material_parameters_updated}. 

The slip resistance due to dislocation hardening $g_{for}^\alpha$ is represented using a Taylor-type hardening model \cite{pan_1982} as:

\begin{equation}
    g_a^\alpha = G b \sqrt{\sum_{\xi=1}^{N_s} A^{\alpha\xi} \rho_{for}^\xi}
\end{equation}

where $G$ is the shear modulus, $b$ is the Burgers vector magnitude, $A^{\alpha\xi}$ is the interaction matrix representing strength of interaction between dislocations on slip systems $\alpha$ and $\xi$ as proposed by \cite{arsenlis2002modeling}, and $\rho_{for}^\xi$ is the forest dislocation density on slip system $\xi$. 
   

 The forest dislocation density is evolved using the phenomenological storage-recovery model proposed by \cite{kocks1976laws} as: 

\begin{equation}
    \frac{\partial \rho_{for}^{\alpha}}{\partial \gamma^{\alpha}} = k_{1}^{\alpha} \sqrt{\rho_{for}^{\alpha}} - k_{2}^{\alpha} (\dot{\varepsilon}, T) \rho_{for}^{\alpha}
\end{equation}

where $\gamma^\alpha(t)=\int_0^t |\dot{\gamma}^\alpha| dt$ is the accumulated plastic strain on slip-system $\alpha$ and the slip-rate $\dot{\gamma}^\alpha$ is given in Eq. \eqref{eq:sliprate}. In the above expression, $k_1$ and $k_2$ are parameters associated with dislocation multiplication and annihilation respectively. The parameter $k_1$ is treated as a constant while $k_2$ is dependent on temperature and strain rate and is given as \cite{essmann1979annihilation}: 

 \begin{equation}
    \frac{k_{2}^{\alpha}}{k_{1}^{\alpha}} = \frac{\bar{\chi} b}{Q} \left( 1 - \frac{kT}{D^{\alpha} b^{3}} \ln \left( \frac{\dot{\varepsilon}}{\dot{\varepsilon}_{0}} \right) \right)
\end{equation}
In this expression, $\bar{\chi}$ denotes the interaction parameter, which is assumed to be 0.9 \cite{zecevic2015dislocation}. The quantities $D, Q, k$, and $\dot{\varepsilon}$ correspond to the drag stress, effective activation enthalpy, Boltzmann constant, and strain rate, respectively. The reference strain rate $\dot{\varepsilon}_0$ is taken as $10^{7} \, \text{s}^{-1}$ \cite{zecevic2015dislocation}.

The back stress $\chi^{\alpha}$ evolves according to
an Armstrong-Fredrick type equation \cite{frederick2007mathematical,mcginty2001multiscale} and is given as: 

\begin{equation}
\dot{\chi}^\alpha = C_1\dot{\gamma}^\alpha - C_2\chi^\alpha|\dot{\gamma}^\alpha|
\end{equation}

where the material constant $C_1$ is the hardening coefficient and $C_2$ is the dynamic recovery coefficient. 


The dynamic recovery constant $C_2$ is modified to incorporate microscopic physical mechanisms like dislocation density and precipitate volume fraction, following the work of \cite{shenoy2008microstructure} and \cite{lu2023crystal}, as:
\begin{equation}
    C_{2} = \frac{\eta_{0} f \frac{Z_{1}}{b \lambda}}{\frac{Z_{1}}{b \lambda} + Z_{2} \sqrt{\sum_{\alpha=1}^{12} \rho_{for}^{\alpha}}}
    \label{eq:C2_dislocation}
\end{equation}
where $b, \lambda, f$ and $\rho_{for}^{\alpha}$ are Burgers vector, spacing of precipitates, volume fraction of precipitates and forest dislocation density, respectively while $\eta_{0}, Z_{1}$ and $Z_{2}$ are material constants. 

To capture the time-dependent deformation and stress relaxation in the creep-fatigue interaction test, a static recovery term ($C_3\chi^\alpha$) is added to the backstress evolution given by the A-F kinematic hardening as follows \cite{lu2023crystal}: 
\begin{equation}
\dot{\chi}^{\alpha} = C_{1} \dot{\gamma}^{\alpha} - C_{2} \chi^{\alpha} |\dot{\gamma}^{\alpha}| + C_{3} \chi^{\alpha}
\end{equation}
The static recovery coefficients are defined by the following expressions,
\begin{equation}
C_{3} = r_{0} [\varphi_{s} + (1 - \varphi_{s}) e^{C_{4}}]
 \label{eq:C3_dislocation}
\end{equation}

\begin{equation}
C_{4} = -\frac{\sum_{\alpha=1}^{12} \rho_{for}^{\alpha}}{\rho_{r}}
\end{equation}

Here, $C_3$ is specified as an exponential function of the total forest dislocation density, $r_{0}$ and $\phi_{s}$ function as material constants that determine the extent of static recovery, while $\rho_{r}$ denotes a reference dislocation density that regulates the rate of stress relaxation across subsequent cycles.

    
\subsection{Creep-Fatigue Life prediction model}
\label{sec:life_model}
Some empirical life prediction models, like frequency-modified \cite{coffin1969predictive} or \cite{coffin1976concept} equations, aren't very good because they only use direct parameter correlation in place of physical understanding. According to Continuum Damage Mechanics (CDM) failure is a process of gradual degradation that can be measured by an internal damage variable, which was first suggested by \cite{kachanov1999rupture}. However, a more fundamental descriptor of this irreversible degradation can be found in thermodynamics. Therefore, precisely describing the damage state is the critical aspect for making a damage-based life prediction model. The Second Law of Thermodynamics dictates that all real-world physical processes are irreversible, resulting in the continuous production of entropy. Consequently, thermodynamic entropy tells us about the the degree of irreversibility in a system. Since material degradation and fatigue are inherently irreversible processes that increase microstructural disorder, it is physically grounded to utilize entropy generation as a direct indicator of damage evolution. Early investigations by \cite{italyantsev1984thermodynamic} established the theoretical basis for this approach, deriving the entropy generation functions (in both integral and differential forms) applicable to deforming solids. This work used the concept of Fracture Fatigue Entropy (FFE) which has been discussed in \cite{naderi2010experimental}—defined as the accumulated entropy at the point of failure which is solely determined by the material and independent of geometry and load amplitude and frequency. Furthermore, a linear relationship has been observed between the normalized entropy and the fatigue cycle ratio. Building on this foundation, \cite{amiri2014entropy} developed a comprehensive thermodynamic framework that employs entropy generation as the primary damage surrogate. This method has been successfully validated for predicting the fatigue life of metals, such as low-carbon steel \cite{idris2018need}. Based on the Clausius-Duhem inequality, the total entropy generation rate by neglecting the entropy generation caused by internal variables can be expressed as follows:

\begin{equation}
    \dot{S} = \frac{W_p}{T} - \frac{1}{T^2} \mathbf{q} \cdot \nabla T,
    \label{eq:entropy_simplified}
\end{equation}
where $W_p$ is the plastic dissipation and the second term is the entropy generation due to heat conduction, $\mathbf{q}$ is  heat flux, $T$ is the working temperature and $\nabla T$ is the temperature gradient. For metallic materials, the stored energy term associated with internal variables is typically negligible, as stated in \cite{naderi2010experimental}.
In the crystal plasticity framework, cyclic entropy generation is primarily governed by dislocation slip~\cite{staroselsky2011creep}. Assuming negligible temperature gradient ($\nabla T \approx 0$) during low-cycle and isothermal high-temperature fatigue~\cite{salimi2019metal}, the heat conduction term vanishes. Consequently, the accumulated entropy per cycle simplifies to the time-integration of plastic dissipation divided by temperature, over all 12 slip systems:
\begin{equation}
    \Delta S = \int_{\text{cycle}} \frac{W_p}{T} = \int_{\text{cycle}} \frac{1}{T} \left[ \sum_{\alpha=1}^{12} (\tau^\alpha - \chi^\alpha) \dot{\gamma}^\alpha \right] dt
    \label{eq:simplified_entropy}
\end{equation}
Building on the static toughness degradation model~\cite{ye2001new}, \cite{NADERI2010875} derived a thermodynamic fatigue damage variable by correlating normalized entropy generation with the load cycle ratio:

\begin{align}
    D_f &= D_0 + (D_{f,c} - D_0) \frac{\ln(1 - S/S_g)}{\ln(1 - S_c/S_g)} \label{eq:entropy_damage_for_fatigue}
\end{align}

\noindent where $D_f$, $D_0$, and $D_{f,c}$ represent the current, initial, and critical damage parameters, respectively. The terms $S$, $S_c$, and $S_g$ denote the current, critical, and fracture entropy generation, while $A$ and $B$ are material constants. The model confirms that irreversible thermodynamic entropy is a reliable and precise method for indicating the system degeneration. It gives a good comparison to experimental measurements and is a better way to anticipate fatigue failure than curve-fitting, the detailed discussion is given in \cite{NADERI2010875}.

To capture the creep damage during the material degradation, a thermodynamic approach is adopted for the creep damage component, offering a more rigorous physical basis than traditional elastoplastic models. Adapting the work of \cite{payten2010strain}, we replaced the standard dissipation parameters with thermodynamic entropy generation. The accumulated creep damage during the hold period ($D_c$) is thus formulated as:

\begin{equation}
    D_c = \int_{0}^{t_d} \frac{1}{\varphi} \left[ \dot{S}_{in} \right]^{1 - n_1} dt
    \label{eq:creep_damage}
\end{equation}

\begin{equation}
    \varphi = B_1 \exp\left( \frac{-Q}{RT} \right)
    \label{eq:phi_param}
\end{equation}

\noindent where $\dot{S}_{in}$ is the inelastic entropy generation rate during the hold stage (Eq.~\ref{eq:simplified_entropy}). The parameter $\varphi$ is temperature-dependent, governed by the activation energy $Q$, the universal gas constant $R$, and the material constant $B_1$ which is dependent on strain amplitude ($\varepsilon_a)$ and can be calculated as $B_1 = 2.0428\times10^{3} - 2.1056\times10^{4} \times \varepsilon_a - 1.23\times10^{7} \times \varepsilon_a^2$. 

In this work, a microstructure-sensitive  framework is employed to investigate the creep-fatigue life of nickel-based superalloys by focusing on the stabilized mechanical response. As the computational cost of the Crystal Plasticity Finite Element Method (CPFEM) is very high, a half-life approach is adopted, where the steady-state hysteresis loop is used as the representative state for damage accumulation. The total creep fatigue interaction life ($N_{c-f}$) is determined by dividing the damage into creep ($D_c$) and fatigue ($D_f$) components. Under the linear damage summation (LDS) rule, failure life is expressed as the inverse of the simple sum of damage components and is discussed in \cite{10.1007/978-3-642-86014-0_6},
\begin{equation}
N_{c-f} = \frac{1}{D_c + D_f}
 \label{eq:miner rule}
\end{equation}

To account for the complex interactions between these mechanisms and to provide a more conservative safety margin for engineering applications, a non-linear damage summation (NDS) rule is also implemented. This rule utilizes a power-law relationship given in \cite{skelton2008creep} to better capture the accelerated degradation observed in high-temperature environments.

\begin{equation}
N_{c-f} = \frac{1}{(D_c^q + D_f^q)}
\end{equation}

\begin{table}[h!]
    \centering
    \caption{Material Parameters.}
    \label{tab:material_parameters_updated}
    \small 
    \begin{tabularx}{\textwidth}{l X @{\quad} l l}
        \toprule
        Symbols  & Value & Symbols & Value \\
        \midrule
        $C_{11}$ & 175 GPa & $C_{12}$ & 108.5 GPa \\
        $C_{44}$ & 95 GPa & $C_1$ & $1 \times 10^6$ \\
        $\dot{\gamma}_0$ & 0.03 & $\rho_0$ & $1 \times 10^8 \text{ mm}^{-2}$ \\
        $m$ & 50 & $k_1$ & 25 \\
        $Q_1$ & 0.3 & $\chi$ & 0.9 \\
        $C_2$ & $1 \times 10^6$ & $a$ & 0.02 \\
        $\dot{\varepsilon}_0$ & $10^7$ & $\eta_0$ & $5 \times 10^4$ \\
        $M$ & 3.06 & $Z_2$ & $1 \times 10^3$ \\
        $Z_1$ & 1.0 & $\varphi_s$ & 10.0 \\
        $r_0$ & $-0.36$ & $D_0$ & 0 \\
        $\rho_r$ & $3\times 10^9 \text{ mm}^{-2}$ & $S_g$ & 0.38 mJ/(mm$^3$K) \\
        $D_{f,c}$ & 0.9 & $n_1$ & 0.6 \\
        $S_c$ & $0.9S_g$ & $\Delta F$ & 173.673 KJ/mol \\
        $Q$ & $6.97 \times 10^{-19}$ J/atom & $\xi$ & 25,000 \\
        $R$ & 8.314 J/mol.K & $g_t(0)$ & 150 MPa \\
        \addlinespace[3pt]
        $B_1$ & \multicolumn{3}{l}{$2.04\times10^{3} - 2.11\times10^{4} \varepsilon_a - 1.23\times10^{7} \varepsilon_a^2$} \\ 
        \addlinespace[5pt]
        $D$ & $5\times 10^4 + 5\times 10^6 \exp\left(-\frac{(T-1033)^2}{1500}\right)$ & $f$ & 0.7 \\ 
        \addlinespace[5pt]
        $K$ & $1.3806\times10^{-23}$ J/K & $G$ & 115 GPa\\
        $b$ & 0.253 nm & $\lambda$ & 100 nm \\
        \bottomrule
    \end{tabularx}
\end{table}


\subsection{Calibration of model parameters}


 
 The elastic constants ($C_{11}$, $C_{12}$, $C_{44}$) were obtained by fitting the elastic part of the stresss-strain curve to uniaxial tension test data in \cite{Zhang2020} and \cite{lu2022dislocation}. The microstructtural properties such as the precipitate volume fraction ($f$) and spacing ($\lambda$) which has been used in Eq.\eqref{eq:C2_dislocation} is taken from \cite{Xiong2015}. 

Because it is not possible to directly measure the parameters that control the evolution of dislocation density ($\rho_0, k_1, \chi, Q_1, D, \xi, \dot{\epsilon}_0$, $s_{t}(0)$, $\Delta F$), they were determined by fitting the simulated response to the  experimental tension test data in \cite{lu2022dislocation} through trial and error. In the same way, a modified A-F framework was used to represent the kinematic hardening behaviour. The parameters used in the kinematic hardening model - which are the dynamic recovery parameters ($\eta_0, Z_1, Z_2$) of Eq.\eqref{eq:C2_dislocation} and static recovery parameters ($r_0, \phi_s, \rho_r$) of Eq.\eqref{eq:C3_dislocation} were determined by calibrating the model response with the results from cyclic hardening and stress relaxation tests from \cite{SHI201331} and \cite{Jing2013}. 

The parameters involved in the damage evolution and life prediction are the initial or pre-existing damage $D_0$, which was set to zero, which means there is no preexisting damage in the material. The critical damage parameter $D_{f,c}$ is set to $0.9$ following  \cite{naderi2010experimental} and is used in Eq.\eqref{eq:entropy_damage_for_fatigue}. The critical entropy generation $S_c$ is treated as the 90\% of the total entropy generation $S_g$ at fracture, which leads to a relationship $S_c=0.9 S_g$. The total entropy parameter $S_g$ and the constants $B_1$ and $n_1$ used in creep damage Eq.\eqref{eq:creep_damage} were obtained by fitting the model prediction to the experimental data of strain-controlled fatigue and creep-fatigue tests, as discussed later in section \ref{sec:cfi_life}.

\section{Results and Discussion}
\label{sec:results}

\subsection{Material Model Validation}

The constitutive model described in Section \ref{sec:method} was implemented using the open-source Finite Element library MOOSE (Multiphysics Object-Oriented Simulation Environment) ~\cite{permann2020moose} to investigate the creep-fatigue interaction behavior of the DD6 superalloy. We used MOOSE because it has advanced finite element capabilities and can handle strongly coupled multiphysics problems. The derivation of Crystal Plasticity Material update follows the work of \cite{McGinty2001}, but with   different evolution equations for internal variables. The code has been developed using the finite deformation FEM framework of $\rho-\text{CP}$, developed in \cite{PATRA2023112182}. A custom material subroutine was developed to incorporate the stress update and material jacobian specific to our model. We used the PETSc (Portable, Extensible Toolkit for Scientific Computation) tool to perform the linear algebra calculations. Adaptive timestepping has been utilized in our work, which critical at the onset of hold periods, where the rapid stress relaxation kinetics required extremely small time steps to maintain accuracy, whereas larger steps were permitted during the stable ramp phases to optimize computational efficiency.

The simulation domain was a cubic representative volume element (RVE) of size of $1 \, \mu\text{m} \times 1 \, \mu\text{m} \times 1 \, \mu\text{m}$ that was divided into a 10×10×10 mesh of hexahedral (HEX8) elements. A mesh convergence analysis was used to choose this particular discretization. 

\begin{figure}[h!]
    \centering
    \includegraphics[width=0.7\textwidth]{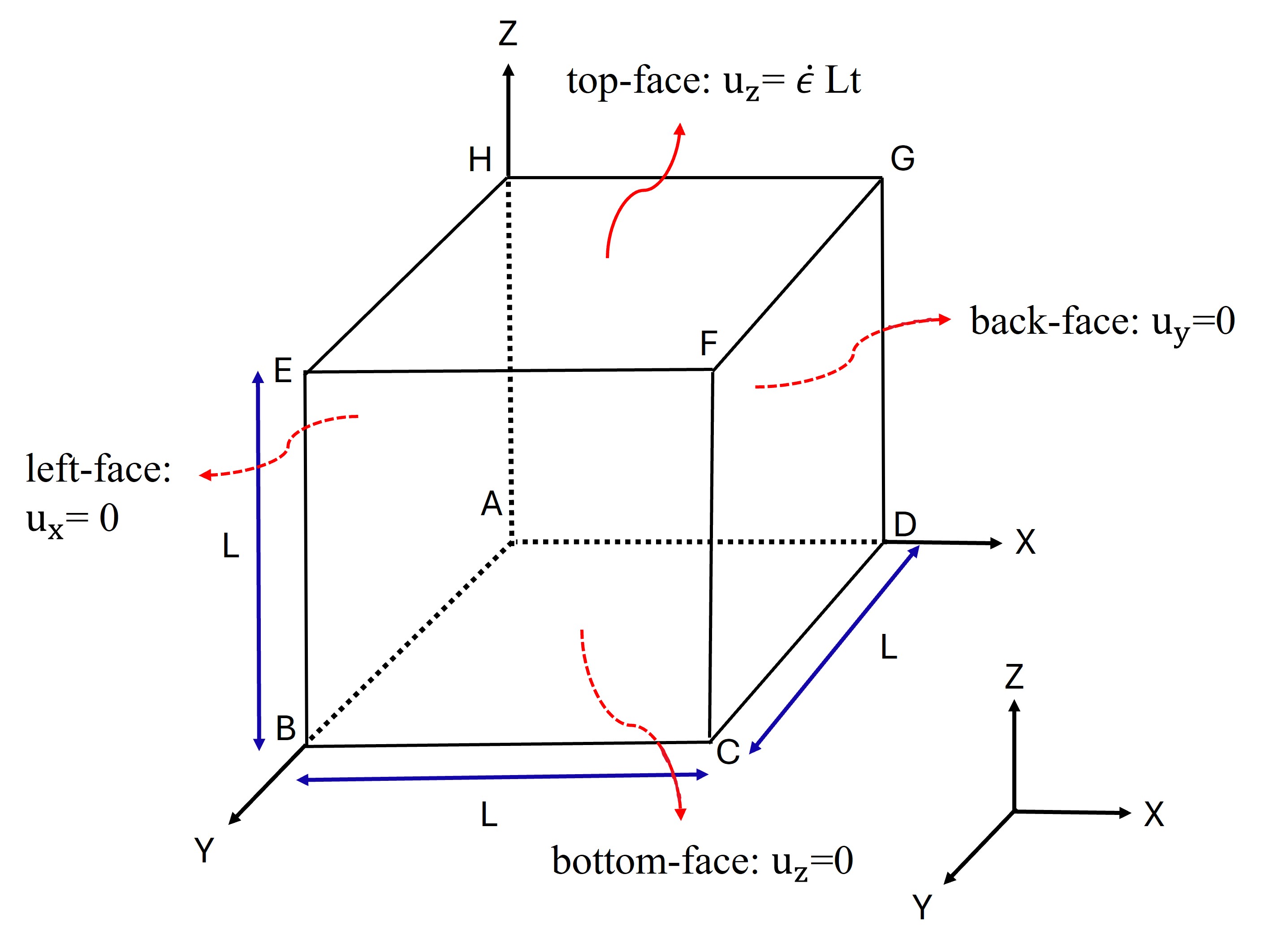}
    \caption{Illustration of the applied boundary conditions.}
    \label{fig:BCs}
\end{figure}

To simulate uniaxial tension test, displacement-controlled boundary condition was applied to the RVE as illustrated in \cref{fig:BCs}. The top face (EFGH) was subjected to a prescribed time-dependent displacement in the Z-direction ($u_{z}=\dot\epsilon Lt$) at a quasi-static strain rate of 10$^{-3}$ s$^{-1}$. To prevent rigid body motion, the following constraints were applied: the left face (ABEH) was fixed in the X-direction ($u_{x}$=0), the bottom face (ABCD) was fixed in the Z-direction ($u_{z}$=0), and the back face (ADGH) was fixed in the Y-direction ($u_{y}$=0). The remaining right face (CDGF) and front face (BCFE) were kept traction-free, allowing the material to undergo free lateral contraction in the X and Y directions due to the Poisson effect.

\begin{figure}[h!]
    \centering
    \captionsetup{format=plain, justification=justified, singlelinecheck=false} 
\includegraphics[width=1.0\textwidth]{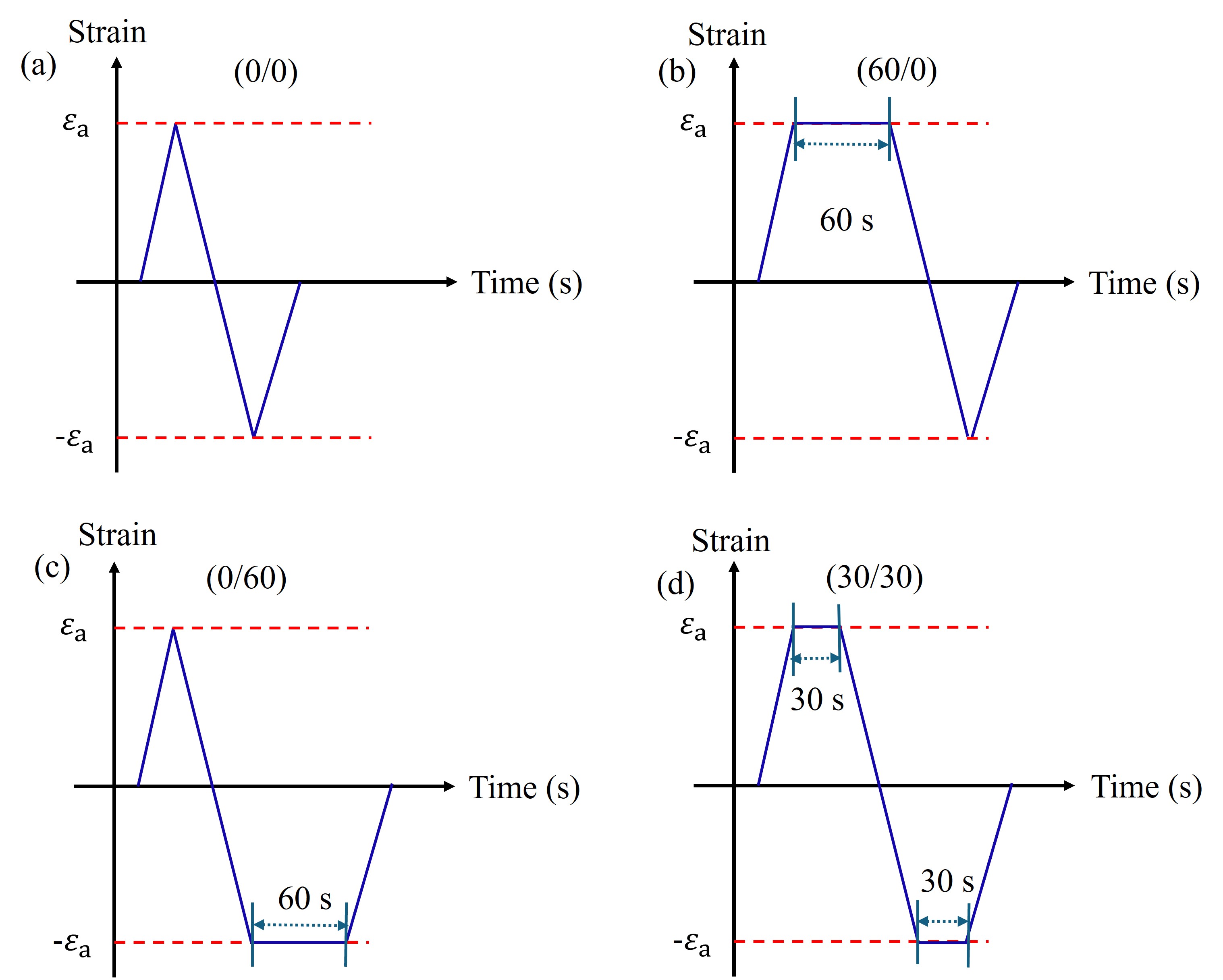}
    \caption{Schematic representation of the applied waveforms for strain-controlled test: (a) pure fatigue with no dwell (0/0); (b) tensile hold only (60/0); (c) compressive hold only (0/60); and (d) combined tensile and compressive holds (30/30).}
    \label{fig:load_history}
\end{figure}
\begin{figure}[h!]
    \centering
    \captionsetup[subfigure]{format=plain, justification=centering, singlelinecheck=false}

    \begin{subfigure}{0.44\textwidth}
        \centering
        \includegraphics[width=\textwidth]{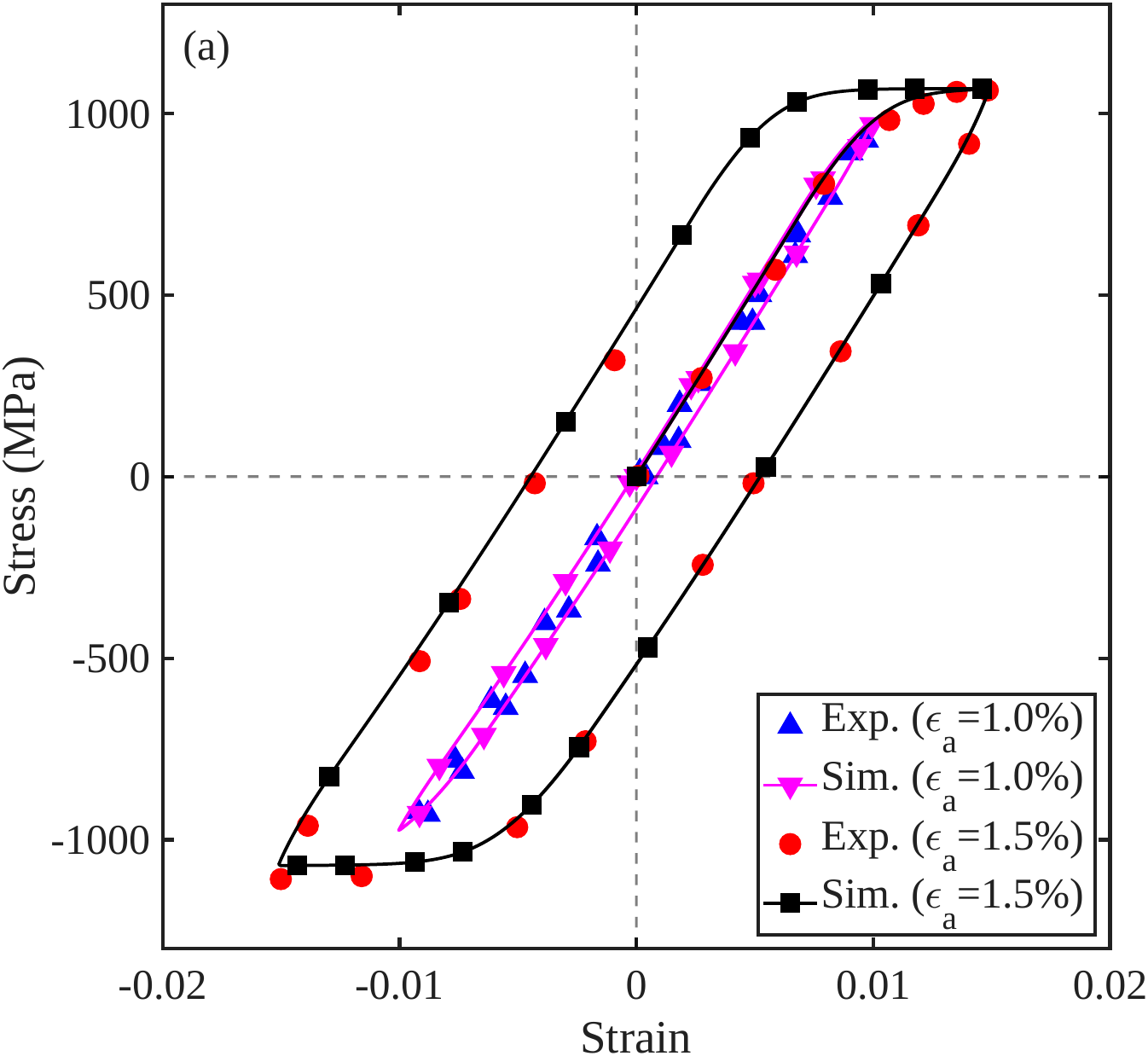}
        \caption{Pure Fatigue (0/0)}
        \label{fig:cyclic_pure_1}
    \end{subfigure}
    \hfill
    \begin{subfigure}{0.44\textwidth}
        \centering
        \includegraphics[width=\textwidth]{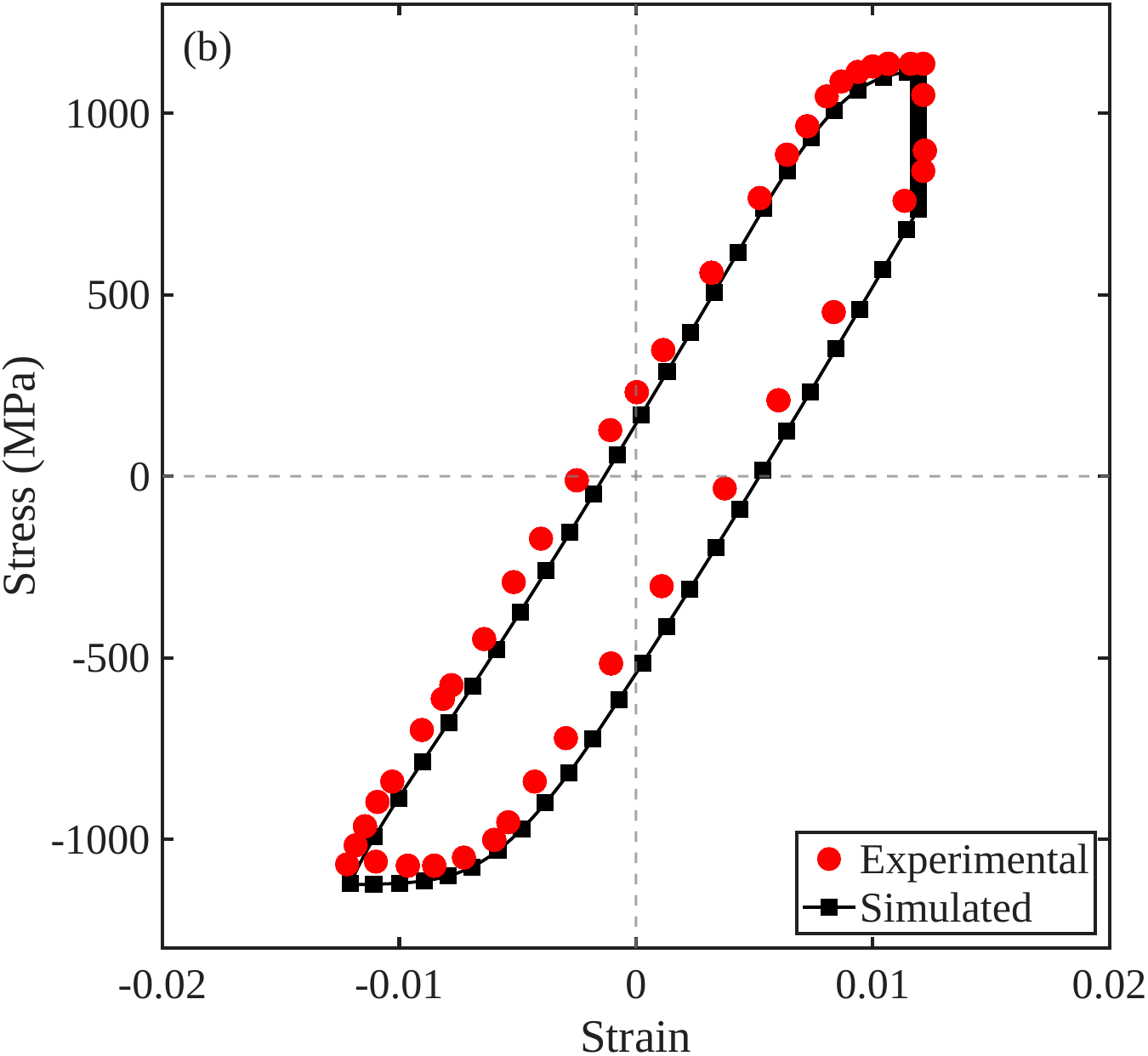}
        \caption{Tensile Dwell (60/0)}
        \label{fig:dwell_tension_1}
    \end{subfigure}
    
    \vspace{0.5cm} 
    
    \begin{subfigure}{0.44\textwidth}
        \centering
        \includegraphics[width=\textwidth]{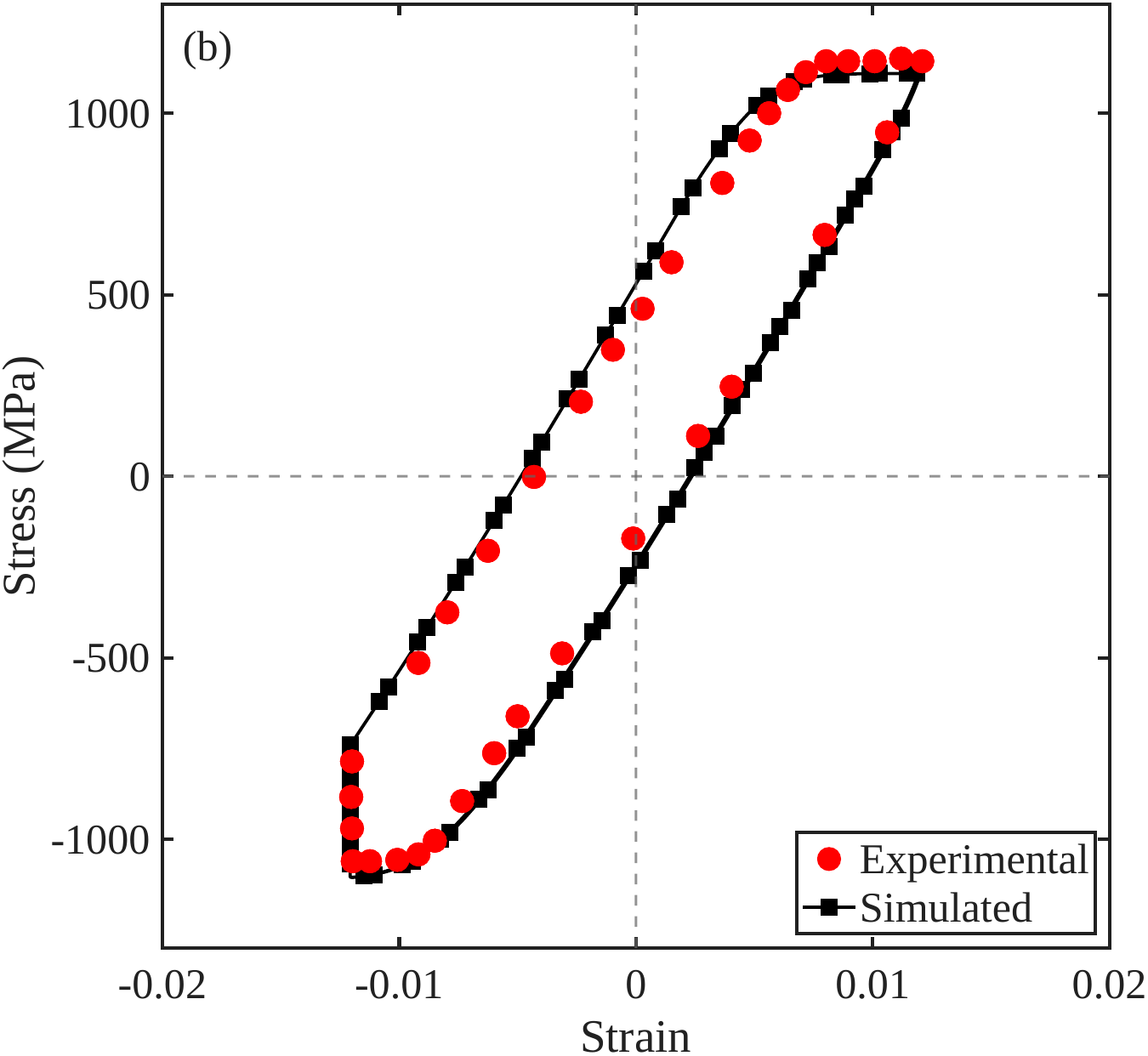}
        \caption{Compressive Dwell (0/60)}
        \label{fig:dwell_compression_1}
    \end{subfigure}
    \hfill
    \begin{subfigure}{0.44\textwidth}
        \centering
        \includegraphics[width=\textwidth]{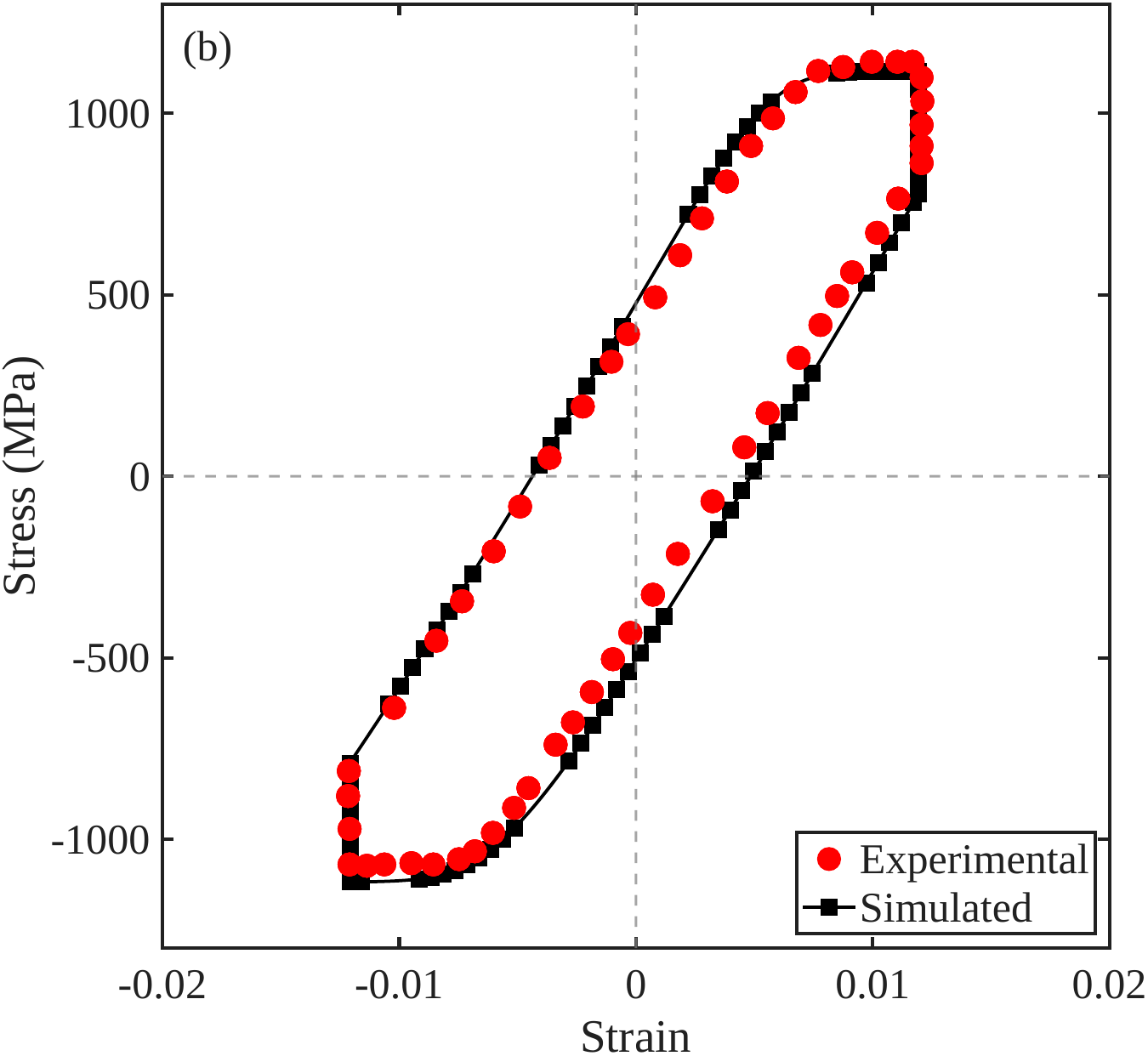}
        \caption{Combined Dwell (30/30)}
        \label{fig:fig1}
    \end{subfigure} 

    \vspace{0.4cm} 
    \captionsetup{justification=justified} 
    \caption{Comparison of experimental results with simulation prediction for cyclic stress–strain hysteresis loops and the experimental data is taken from \cite{lu2023crystal} (a) Pure fatigue response at a strain amplitude of 1.2\%. (b–d) Creep–fatigue response under various holding conditions at 1.2\% strain amplitude: (b) 60 s dwell at peak tension (60/0); (c) 60 s dwell at peak compression (0/60); and (d) symmetric 30 s dwell at both tension and compression peaks (30/30).}
    \label{fig:cyclic_result}
\end{figure}

\begin{figure}[h!]
    \centering
    \includegraphics[width=0.44\textwidth]{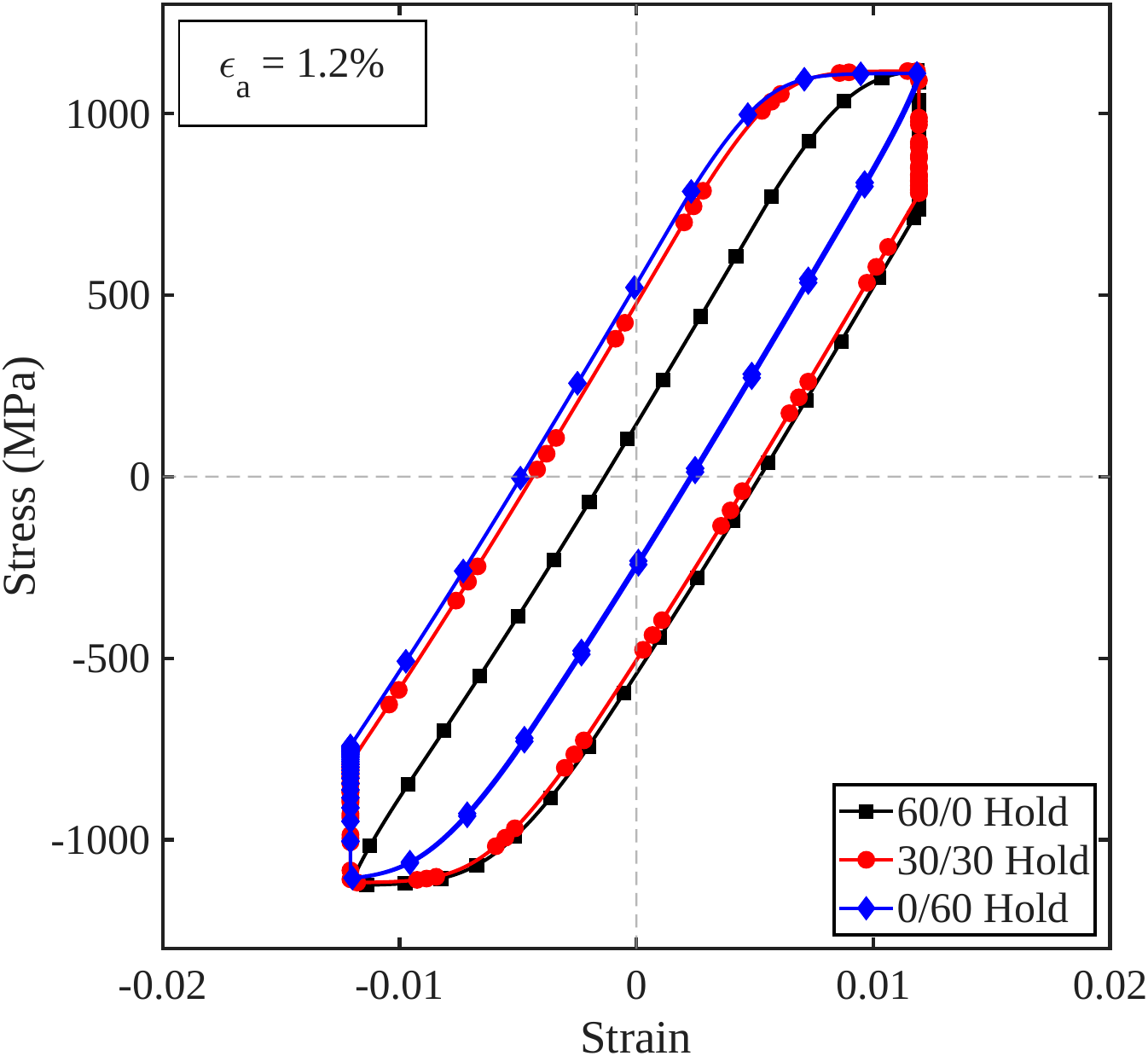}
    \caption{Simulated hysteresis response at 1.2\% strain amplitude. The vertical stress drops at the peak/valley strains represent the predicted stress relaxation during the $60\,s$ tension hold (black), $30\,s$tension/compression hold (red), and $60\,s$ compression hold (blue) cycles.}
    \label{fig:three_loops}
\end{figure}

\begin{figure}[h!] 
    \centering
    \captionsetup[subfigure]{format=plain, justification=centering, singlelinecheck=false}
    
    \begin{subfigure}{0.44\textwidth}
        \centering
        \includegraphics[width=\textwidth]{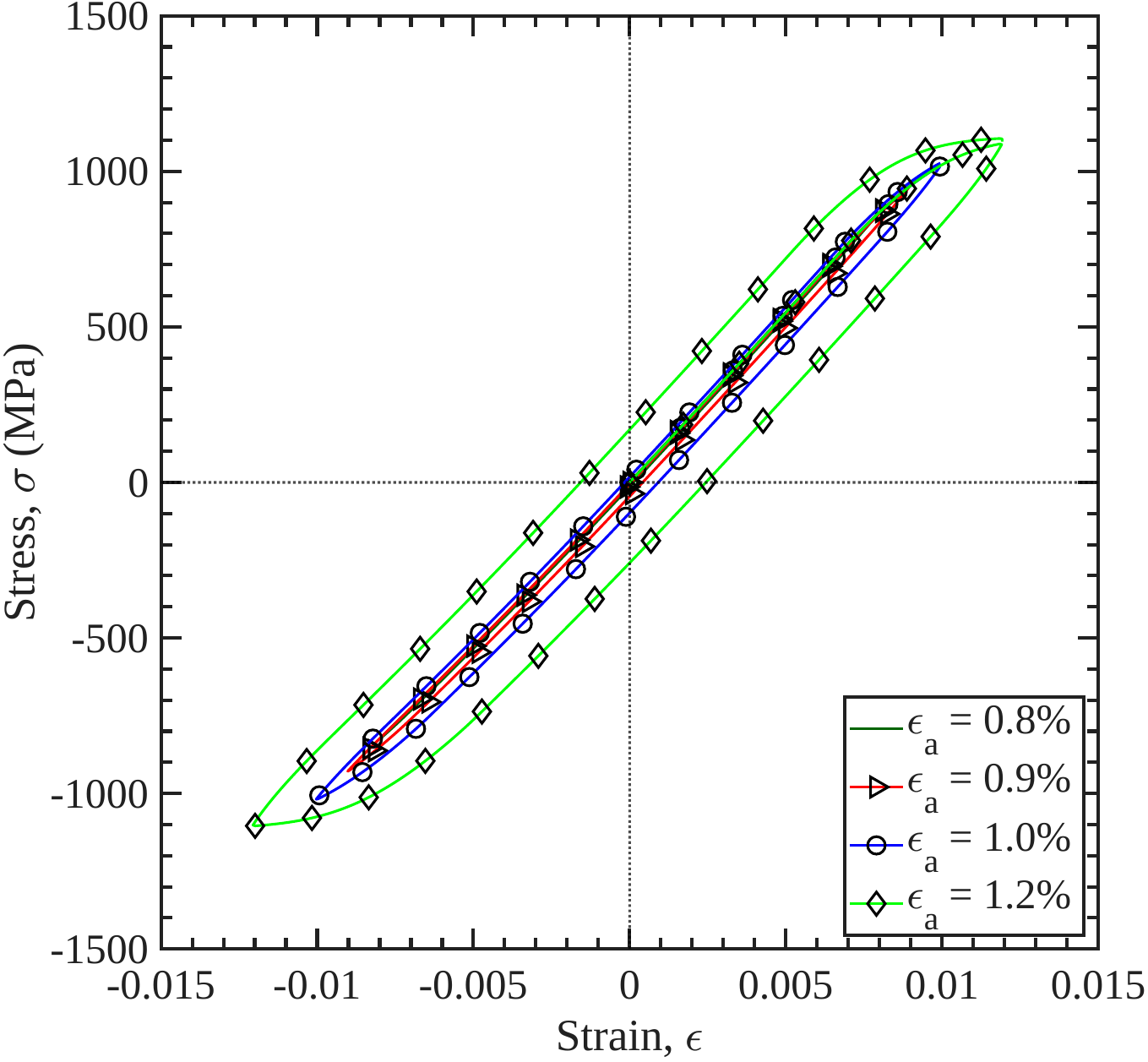}
        \caption{} 
        \label{fig:stress_amp_new}
    \end{subfigure}
    \hfill 
    \begin{subfigure}{0.42\textwidth}
        \centering
        \includegraphics[width=\textwidth]{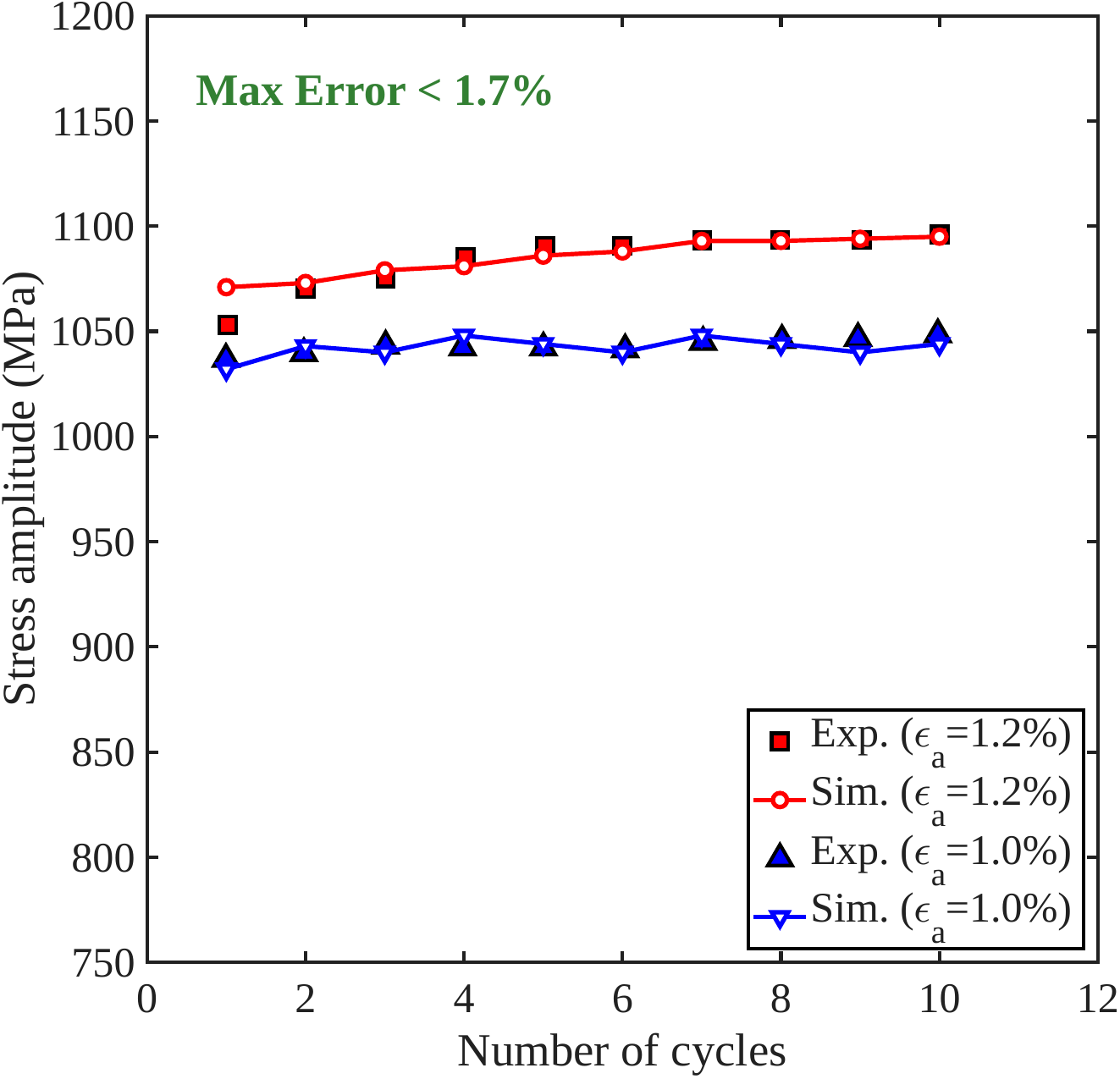}
        \caption{} 
        \label{fig:stress_amp_new_updated}
    \end{subfigure}
    \captionsetup{justification=justified} 
    \caption{Experimental and simulated material response: (a) Stabilized stress-strain hysteresis loops across varying strain amplitudes $(\epsilon_a)$ from 0.8\% to 1.2\% in pure fatigue;  (b) Evolution of stress amplitude over cycles (``Sim.'' denotes simulation predictions while ``Exp.'' denotes experimental data from \cite{lu2023crystal}).}
    \label{fig:fatigue_life_stress_evolution}
\end{figure}

The influence of hold times on the stabilized hysteresis loop was analyzed by implementing the precise strain-time waveforms shown in \cref{fig:load_history}. While the pure fatigue case (0/0) followed a continuous, triangular loading path, the introduction of dwell periods allowed for the critical evaluation of stress relaxation. During these hold periods, the ability of the model to predict the response during the transition from fatigue to creep loading was tested. Specifically, the 60/0 and 0/60 cases subjected the material to $60\,s$ of hold at the tensile and compressive peaks, respectively, providing insight into the asymmetric damage accumulation. In the balanced loading configuration (30/30), a $30\,s$ hold was applied at both the tensile and compressive peaks, as shown in \cref{fig:load_history}d. 

The crystal plasticity constitutive model is validated by fitting the simulation response to experimental results from \cite{SHI201331}, for both cyclic stress-strain hysteresis loops. As shown in \cref{fig:cyclic_result}, the model accurately captures the response under varying loading conditions, including pure fatigue (0/0) and various hold time configurations (60/0, 0/60, and 30/30) at 1.2\% strain amplitude. In \cref{fig:three_loops}, we compare the hysteresis loops for each hold configuration. For tensile hold (60/0), the stress relaxation at the peak tensile strain leads to a downward shift of the loop, resulting in a compressive mean stress which resists fatigue crack growth. But for compressive hold (0/60), the loop shifts upward and tensile mean stress accelerates damage accumulation and promote the nucleation of creep cavities, as justified in \cite{GOYAL201816}  and thereby shortens creep fatigue life. Hold configuration (30/30) shows relaxation at both ends which leads to a balanced but complex interaction of vacancy diffusion and dislocation climb. The \emph{area} inside each loop represents the plastic dissipation per cycle ($W_p$). A larger loop area generally correlates to higher damage per cycle. 
Stabilized stress-strain hysteresis loops (after the $10^{th}$ cycle) are shown in \cref{fig:stress_amp_new} for strain amplitudes ($\epsilon_a=\Delta\epsilon/2$) ranging from 0.8\% to 1.2\% in pure-fatigue conditions (with zero hold time). As ${\epsilon}_a$ increases, the hysteresis loop area also increases. This means that the plastic dissipation ($W_{p}$ as defined in Eq. \eqref{eq:simplified_entropy}) in every cycle also increases. This shows that plastic deformation is promoted at larger amplitudes as expected, which leads to increased acculumation of fatigue-induced inelastic strain.     \cref{fig:stress_amp_new_updated} shows the change in stress amplitude with the first 10 loading cycles under pure fatigue condition. The comparison between experimental data from \cite{lu2023crystal} and our simulation data shows close agreement with a maximum error of less than 1.7\%. The material shows a slight cyclic hardening behavior in the first 10 cycles before stabilizing.  

\subsubsection{Creep-fatigue life prediction}\label{sec:cfi_life}

\begin{figure}[h!] 
    \centering
    \captionsetup[subfigure]{format=plain, justification=centering, singlelinecheck=false}
    
    \begin{subfigure}[b]{0.483\textwidth}
        \centering
         \includegraphics[width=\textwidth]{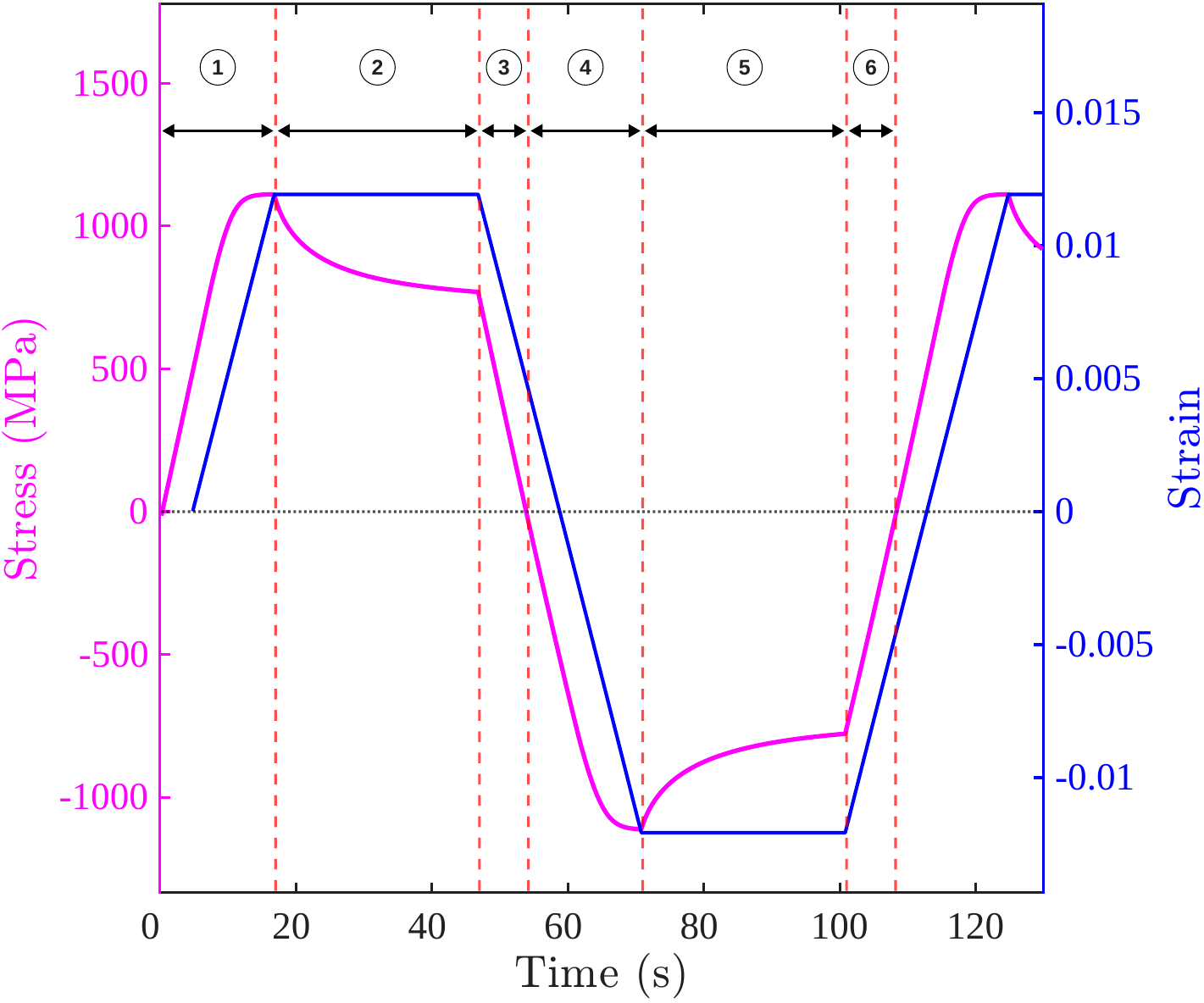}
        \caption{} 
          \label{fig:stress_time_new}
    \end{subfigure}
    \hfill
    \begin{subfigure}[b]{0.465\textwidth}
        \centering
       \includegraphics[width=\textwidth]{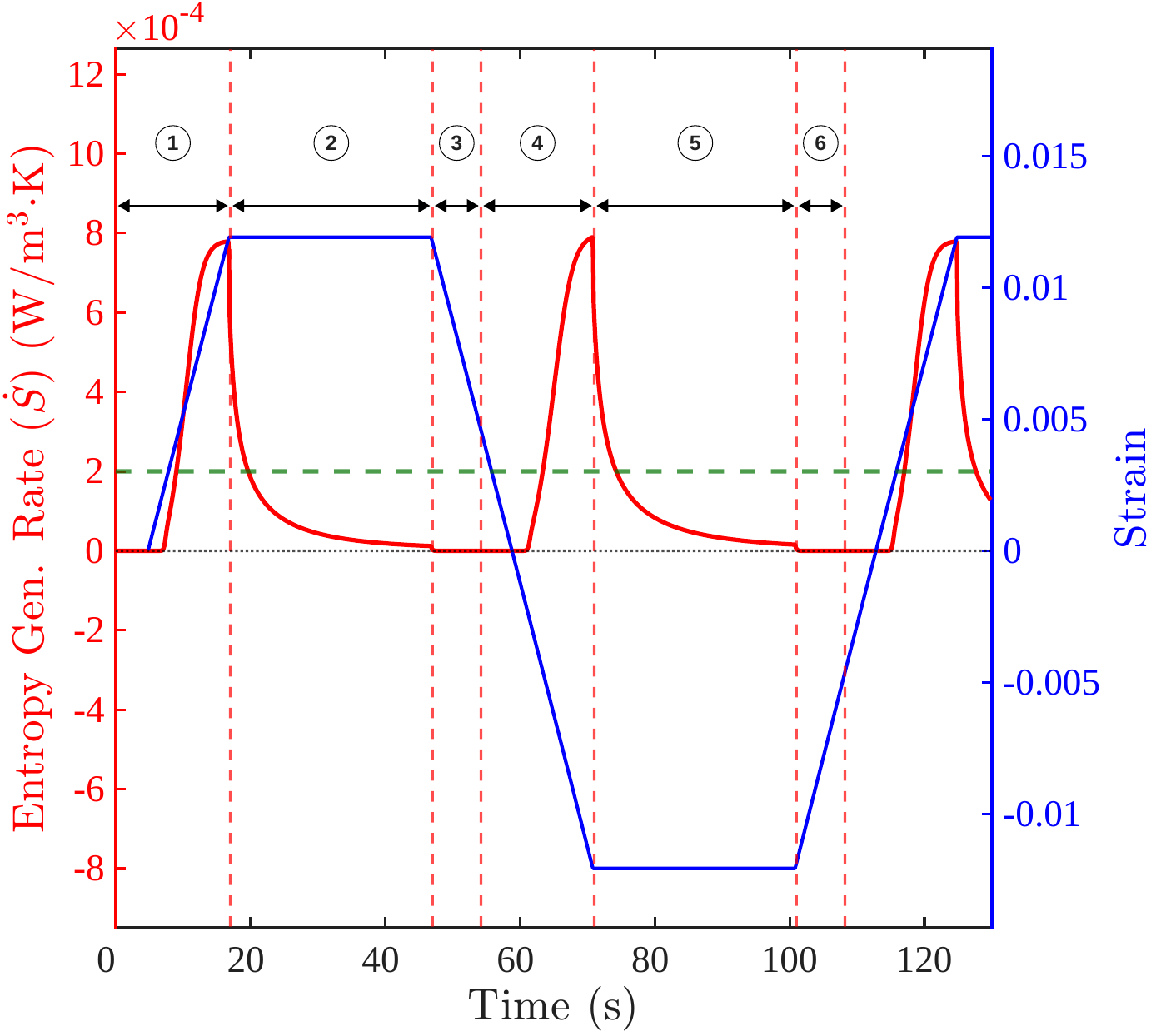}
        \caption{} 
         \label{fig:entroy_generation_new}
    \end{subfigure}
 
    \vspace{0.4cm} 
    \captionsetup{justification=justified} 
    \caption{(a) Evolution of stress and strain with time during a stabilized creep-fatigue cycle. Stress rises during Stage 1 (tensile loading) followed by stress relaxation during Stage 2 (tensile hold). Stress drops during Stage 3 (tensile unloading) and drops further in compression in Stage 4 (compressive loading), followed by relaxation in Stage 5 (compressive hold) and stress rise in Stage 6 (compressive unloading) (b) Evolution of entropy generation rate ($\dot S$) and strain with time during a stabilized creep-fatigue cycle. Fatigue damage occurs during Stage 1 and 4 while creep damage occurs during the early phases of Stage 2 and 5. A horizontal line at $2 \times 10^{-4}$ in green is marked to show the stages of the loading cycle which are associated with high rate of fatigue or creep damage.}
    \label{fig:damage_analysis}
\end{figure}

\cref{fig:stress_time_new} shows the evolution of stress  with time in the $10^{th}$ creep-fatigue cycle, after the hysteresis loop has stabilized under the application of balanced (30/30 s) hold. A single creep-fatigue cycle consists of 6 distinct stages: Stage 1 (tensile loading), Stage 2 (tensile hold), Stage 3 (tensile unloading), Stage 4 (compressive loading), Stage 5 (compressive hold), Stage 6 (compressive unloading). During the tensile loading (Stage 1) and compressive loading (Stage 4), the stress increases rapidly due to hardening caused by dislocation multiplication which causes inelastic fatigue strain. In the hold stages (2 and 5), stress relaxation occurs due to the conversion of elastic strain into inelastic creep strain, which continues till the dislocation network reaches a stable network structure, a phenomenon observed in prior literature (\cite{WANG2020105879}). In Stages 3 and 6, unloading takes place and no damage occurs in these stages. Therefore, Stages 1 and 4 are used to calculate fatigue damage and stages 2 and 5 are used to calculate creep damage in each cycle. The fatigue and creep damages are calculated using  Eq. \eqref{eq:entropy_damage_for_fatigue} and Eq. \eqref{eq:creep_damage}, both of which primarily depend on the entropy generation rate $\dot S$. Hence, the evolution of $\dot{S}$ with time as shown in \cref{fig:entroy_generation_new} gives a measure of the rate of damage at different stages of the creep-fatigue cycle, as discussed next. In Stages 1 and 4 the high rate of entropy generation indicates a high rate of fatigue damage. In Stages 2 and 5, the rate of entropy production is still high particularly at the beginning of the hold period and sharply reduces after that, which indicates creep damage in these stages. There is negligible amount of entropy generation in the stages 3 and 6, hence no damage occurs in these stages as expected.

\begin{figure}[h!]
    \centering
     \captionsetup[subfigure]{format=plain, justification=centering, singlelinecheck=false}
    \begin{subfigure}[b]{0.415\textwidth}
        \centering
        \includegraphics[width=\textwidth]{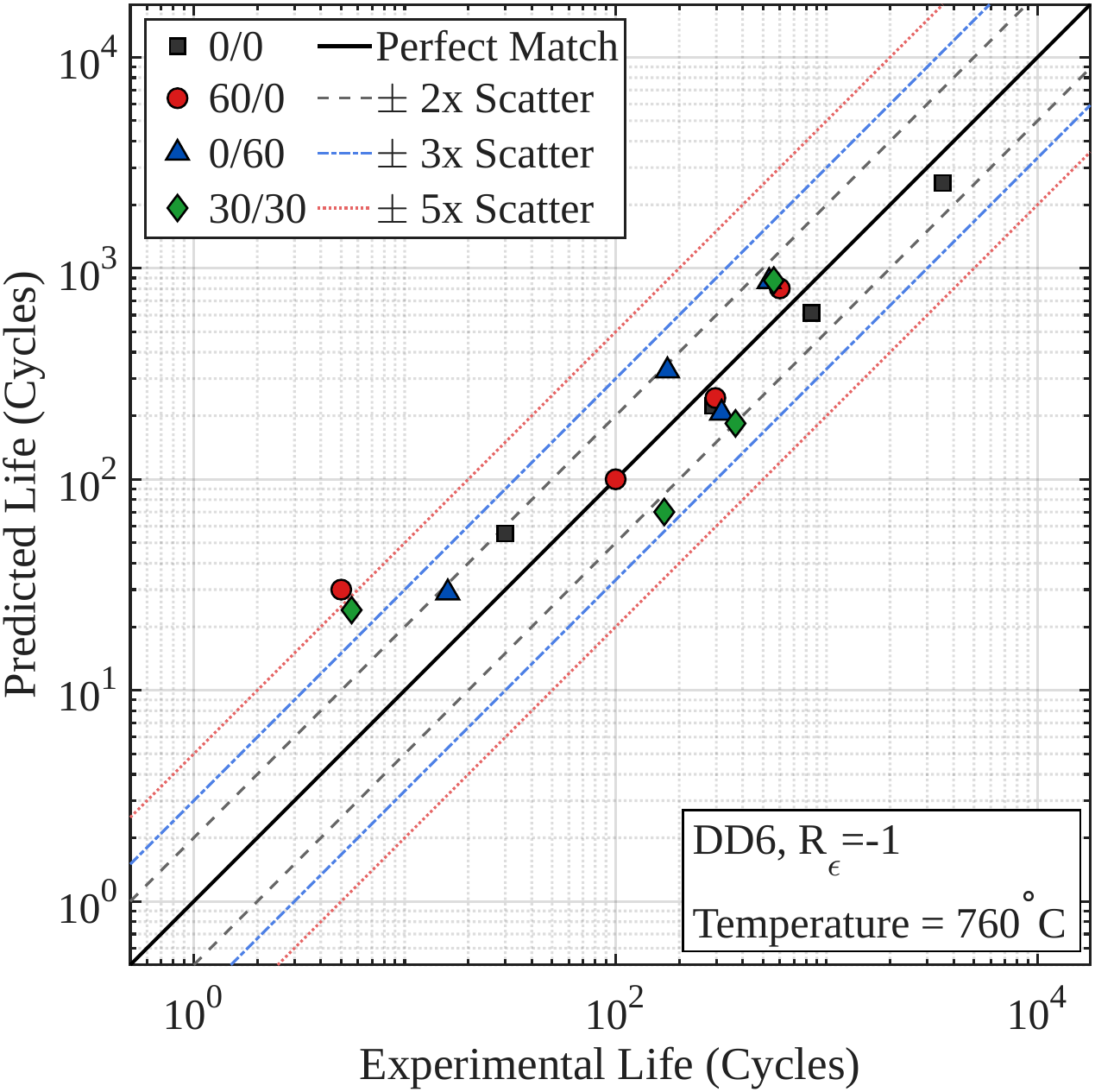}
        \caption{}
        \label{fig:life_760_a}
    \end{subfigure}
    \hfill 
    \begin{subfigure}[b]{0.415\textwidth}
        \centering
        \includegraphics[width=\textwidth]{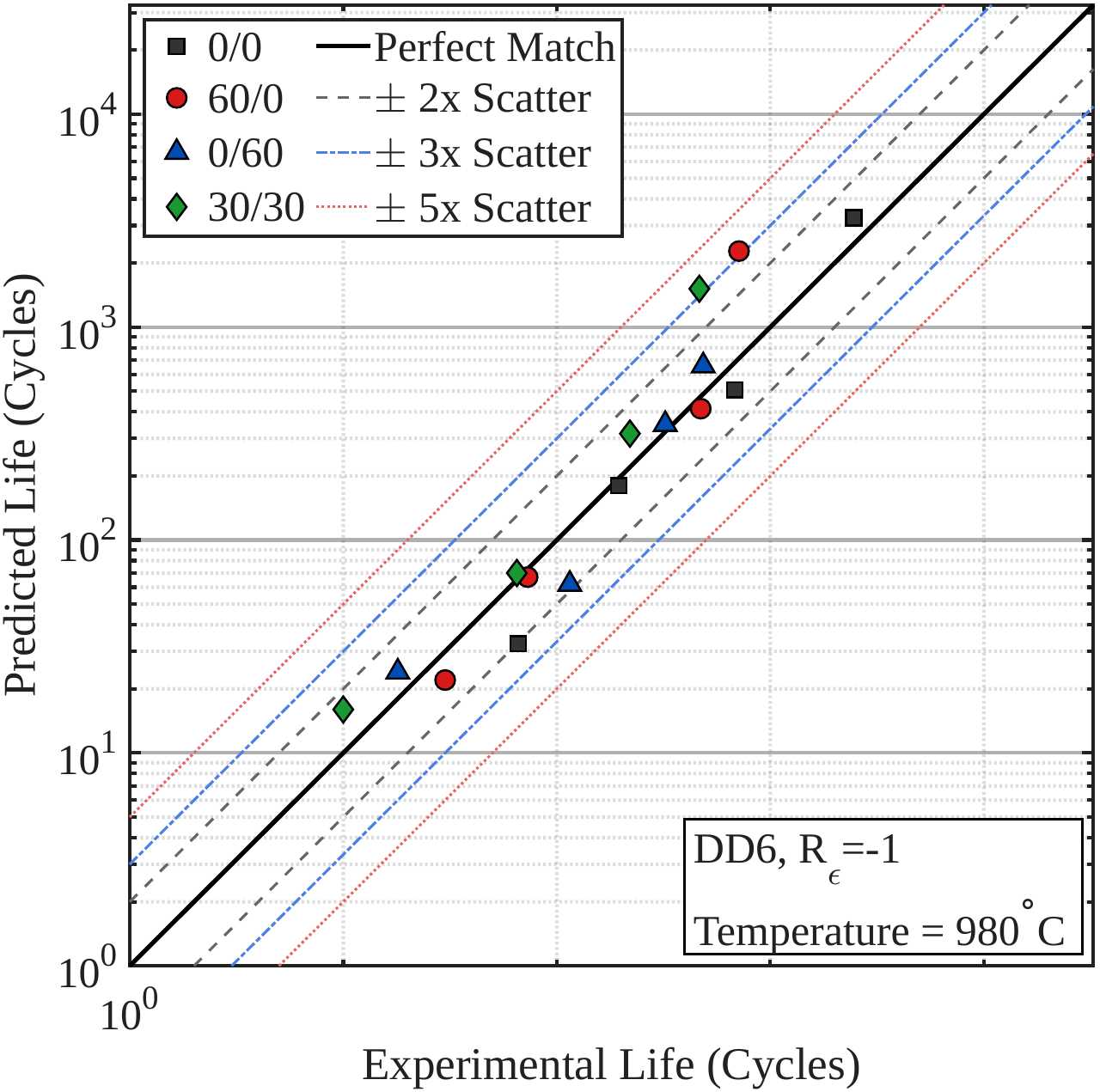}
        \caption{}
        \label{fig:life_980}
    \end{subfigure}
    
    \vspace{0.4cm}
    \captionsetup{justification=justified}
    \caption{Comparison between experimental and predicted creep fatigue life under various hold conditions: (a) 760°C; and (b) 980°C.}
    \label{fig:combined_life_plots_111}
\end{figure}

\begin{figure}[h!]
    \centering
     \captionsetup[subfigure]{format=plain, justification=centering, singlelinecheck=false}
    \begin{subfigure}[b]{0.415\textwidth}
        \centering
        \includegraphics[width=\textwidth]{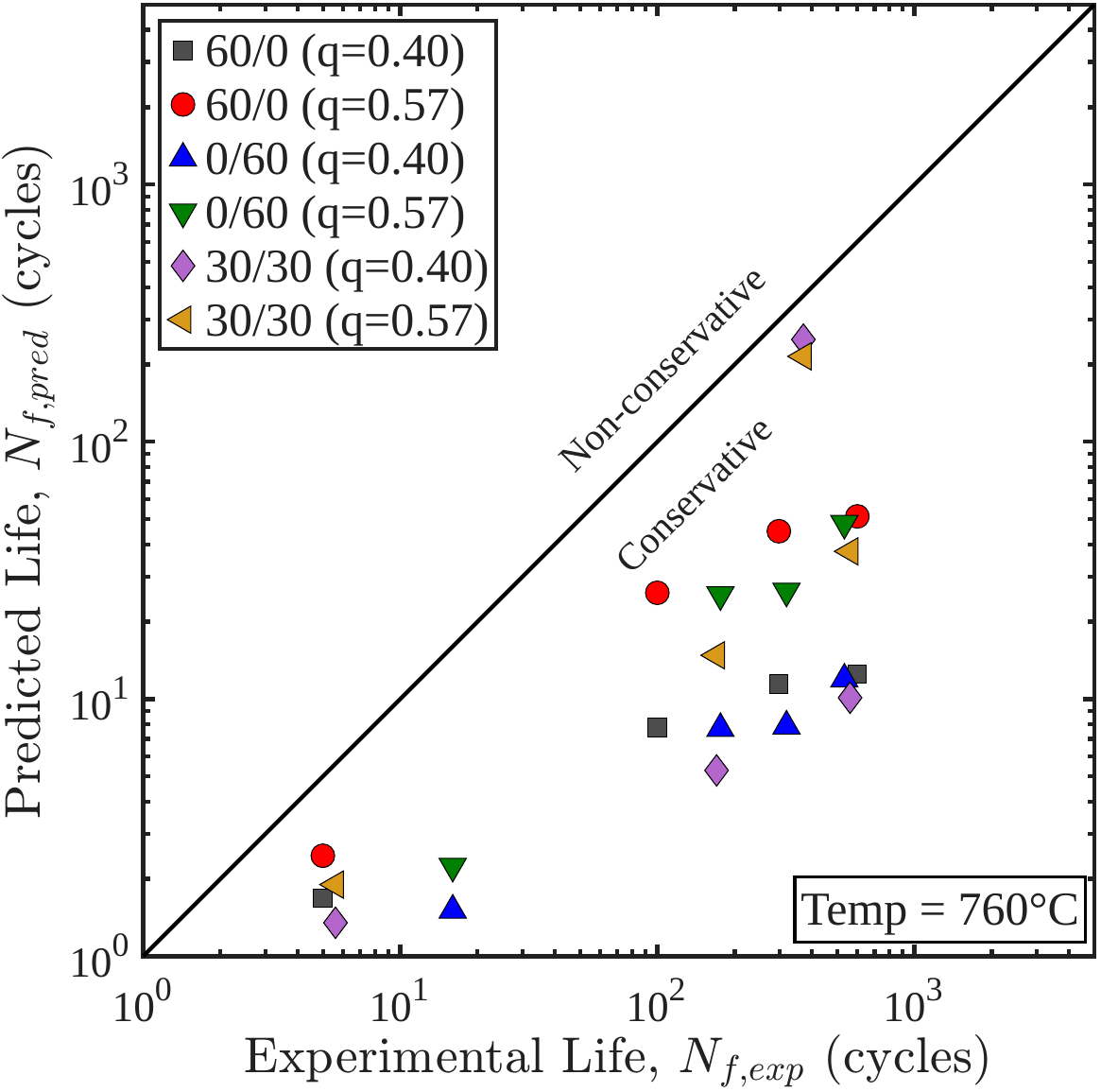}
        \caption{}
        \label{fig:life_760_a}
    \end{subfigure}
    \hfill 
    \begin{subfigure}[b]{0.415\textwidth}
        \centering
        \includegraphics[width=\textwidth]{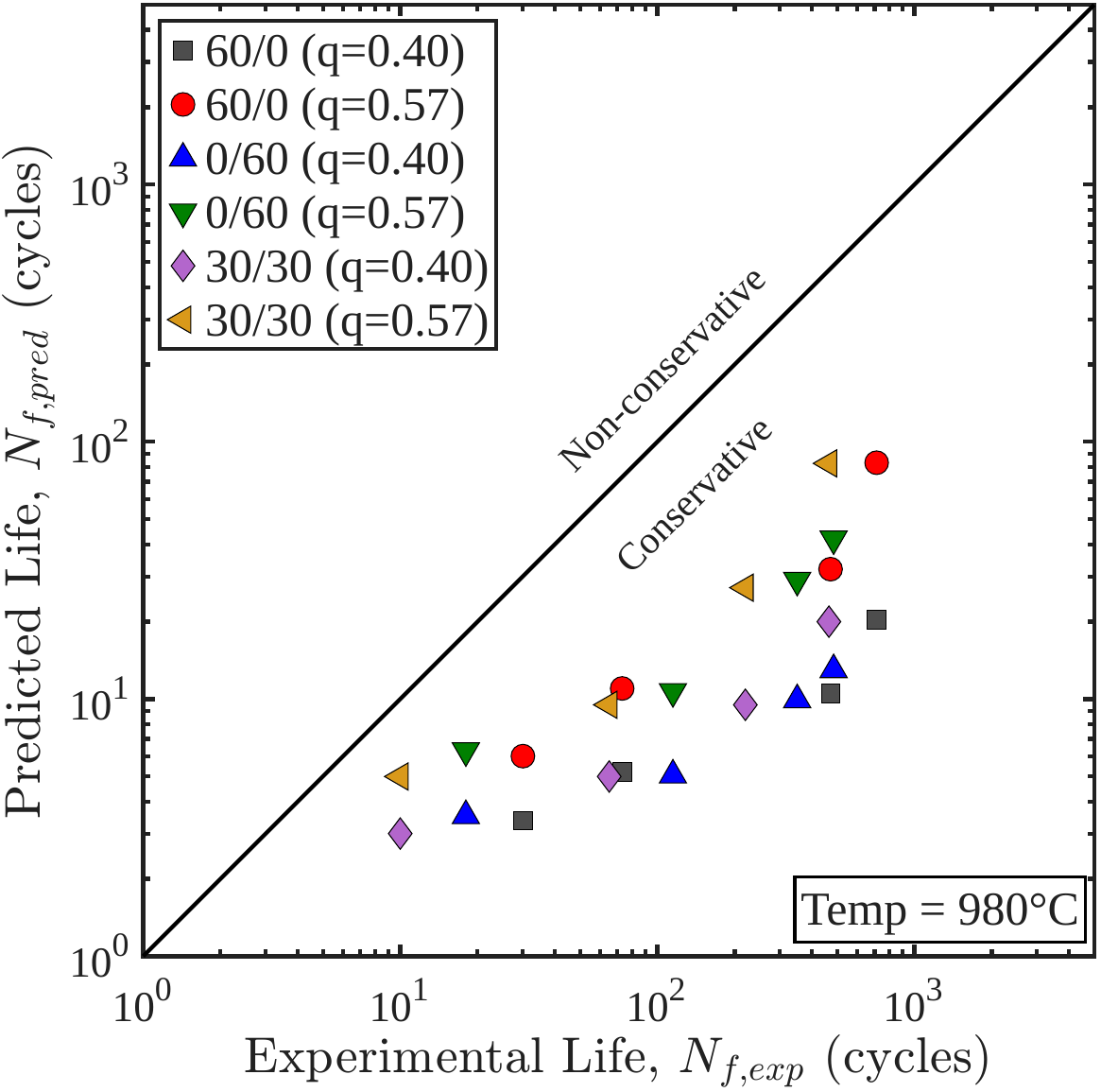}
        \caption{}
        \label{fig:life_980}
    \end{subfigure}
    
    \vspace{0.4cm}
    \captionsetup{justification=justified}
    \caption{Comparisons between experimental and predicted life following the NDS rule under various hold conditions: (a) 760° C and (b) 980° C.}
    \label{fig:combined_life_NDS}
\end{figure}

The coupled damage mechanism under creep-fatigue loading needs a unified explanation. It is known that metallic materials have a dispersive creep-fatigue life distribution. The LDS rule \cref{eq:miner rule} defines material damage as a linear accumulation of creep and fatigue damage. In the absence of more physics based model of CFI, we use the LDS rule as a simplified approximation to determine total damage and life prediction. The correlation between the experimental life and the simulation predictive life is shown in \cref{fig:combined_life_plots_111}. The creep-fatigue life data of tension peak strain hold-period of 60 s with different strain amplitudes were used for model calibration. The experimental data is taken from \cite{Shi2013}. These plots compare the experimental life against the predicted life, where the solid diagonal line represents a \emph{Perfect Match}. For both the 760°C and 980°C test cases, the majority of the data points—representing different hold configurations such as 0/0, 60/0, 0/60, and 30/30—fall within the ±2x scatter bands.

The assessment of the Non-linear Damage Summation (NDS) rule at both 760°C and 980°C are shown in \cref{fig:combined_life_NDS}, shows a significant trend toward conservative predictions when compared to experimental data. Unlike the previous correlation plots, the data points in these plots fall almost below the solid 1:1 \emph{perfect match} line and lie within the region labeled \emph{More conservative results}. This indicates that the NDS rule tends to predict a shorter creep fatigue life than what is actually observed in the experiments. The degree of conservatism is influenced by the exponent parameter $q$, where results calculated with $q=0.576$ generally provide life estimates that are slightly closer to the experimental values than those using q=0.4, though both remain on the safe, conservative side of the design curve.

\section{Parametric analysis of creep-fatigue interaction and life prediction}

\subsection{Effect of hold time and strain amplitude}

\begin{figure}[h!] 
    \centering
    \captionsetup[subfigure]{format=plain, justification=centering, singlelinecheck=false}
    \begin{subfigure}{0.435\textwidth} 
    \centering
    \includegraphics[width=\textwidth]{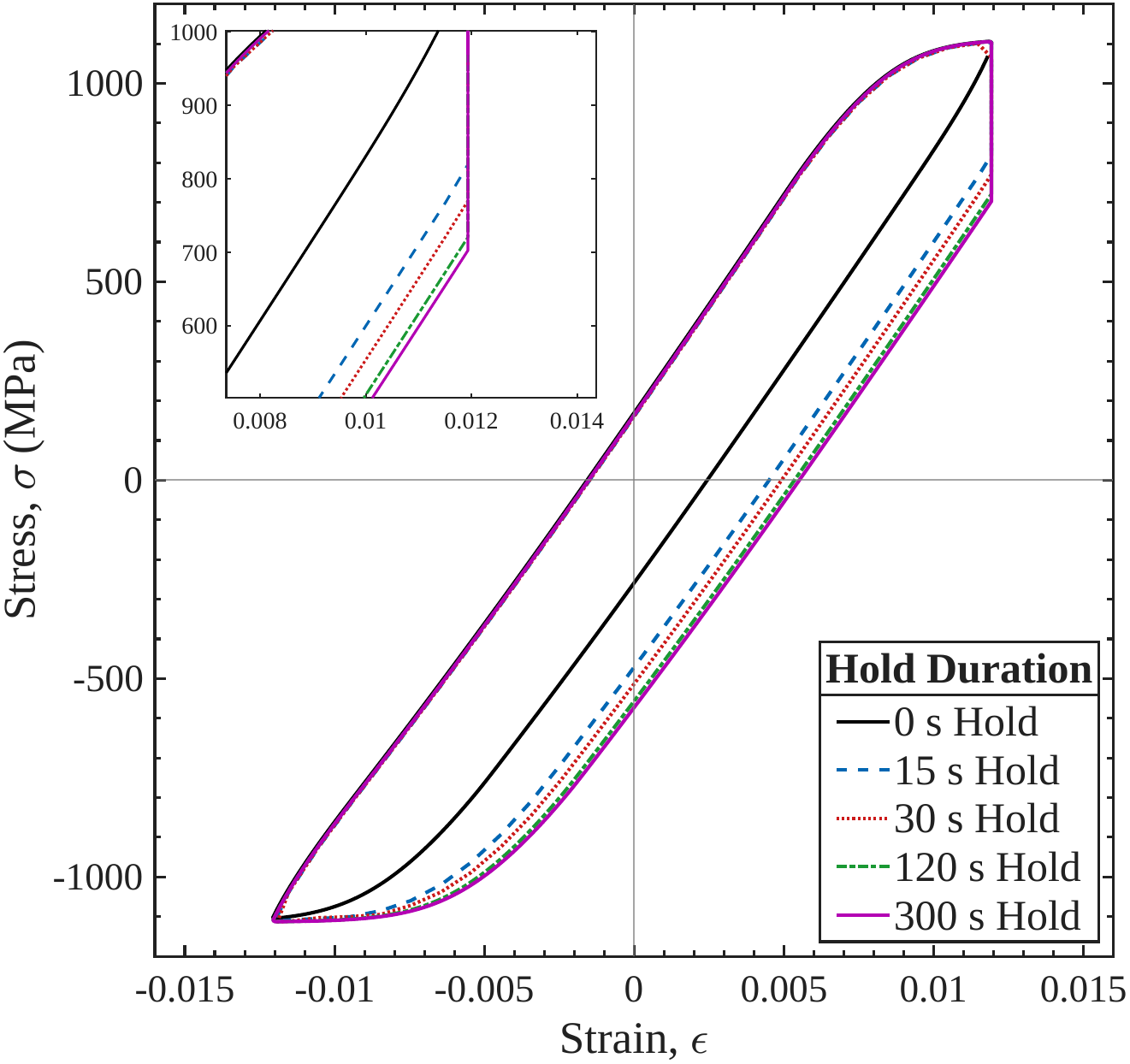}
    \caption{} 
    \label{fig:holdtime}
\end{subfigure}
\hfill 
\begin{subfigure}{0.43\textwidth}
        \centering
        \includegraphics[width=\textwidth]{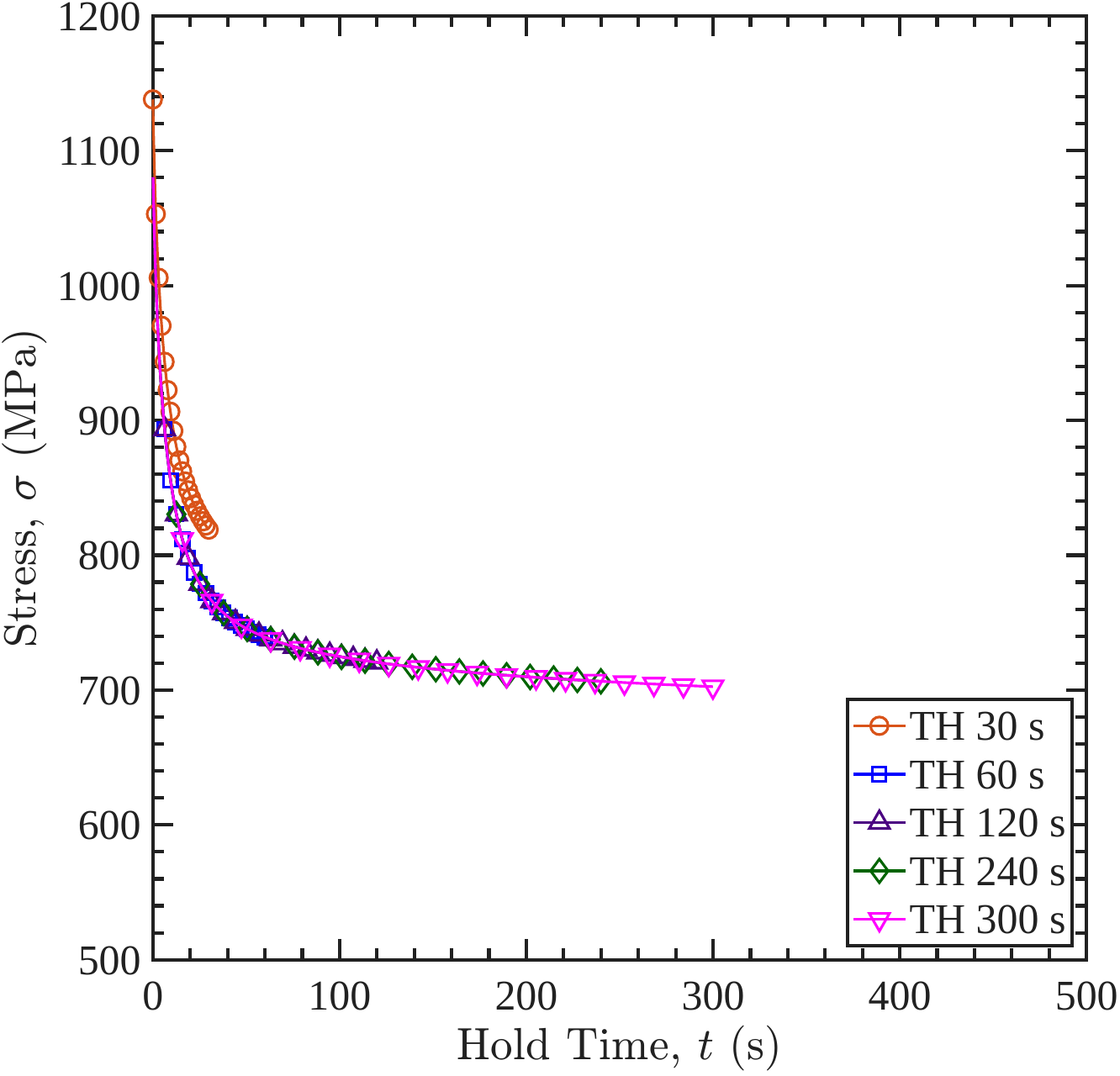}
        \caption{} \label{fig:life_vs_mean_2}
    \end{subfigure}
    
    \vspace{0.5cm} 
    
    \begin{subfigure}{0.424\textwidth}
        \centering
        \includegraphics[width=\textwidth]{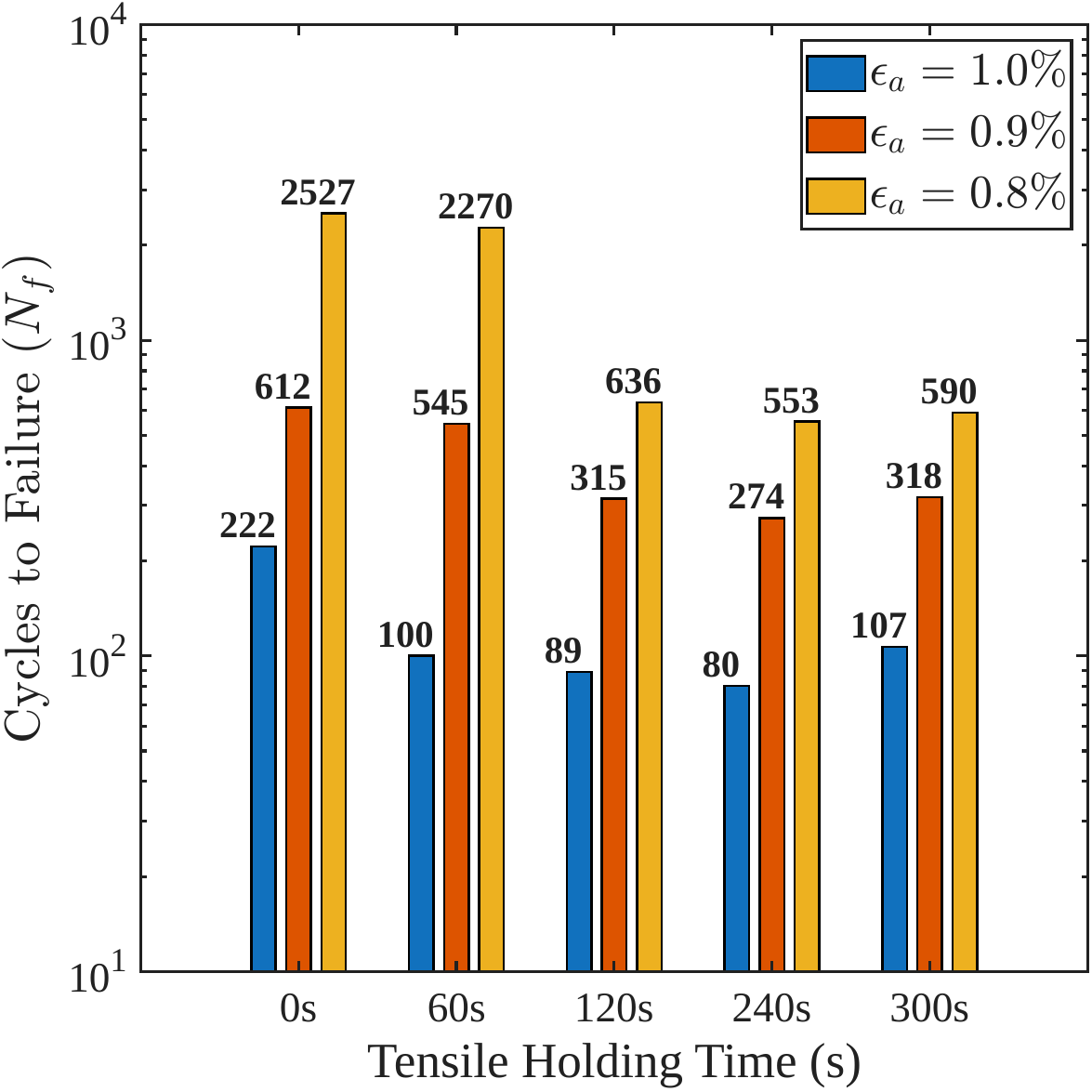}
        \caption{} \label{fig:life_vs_ratio_2}
    \end{subfigure}
    \hfill
    \begin{subfigure}{0.417\textwidth} 
    \centering
    \includegraphics[width=\textwidth]{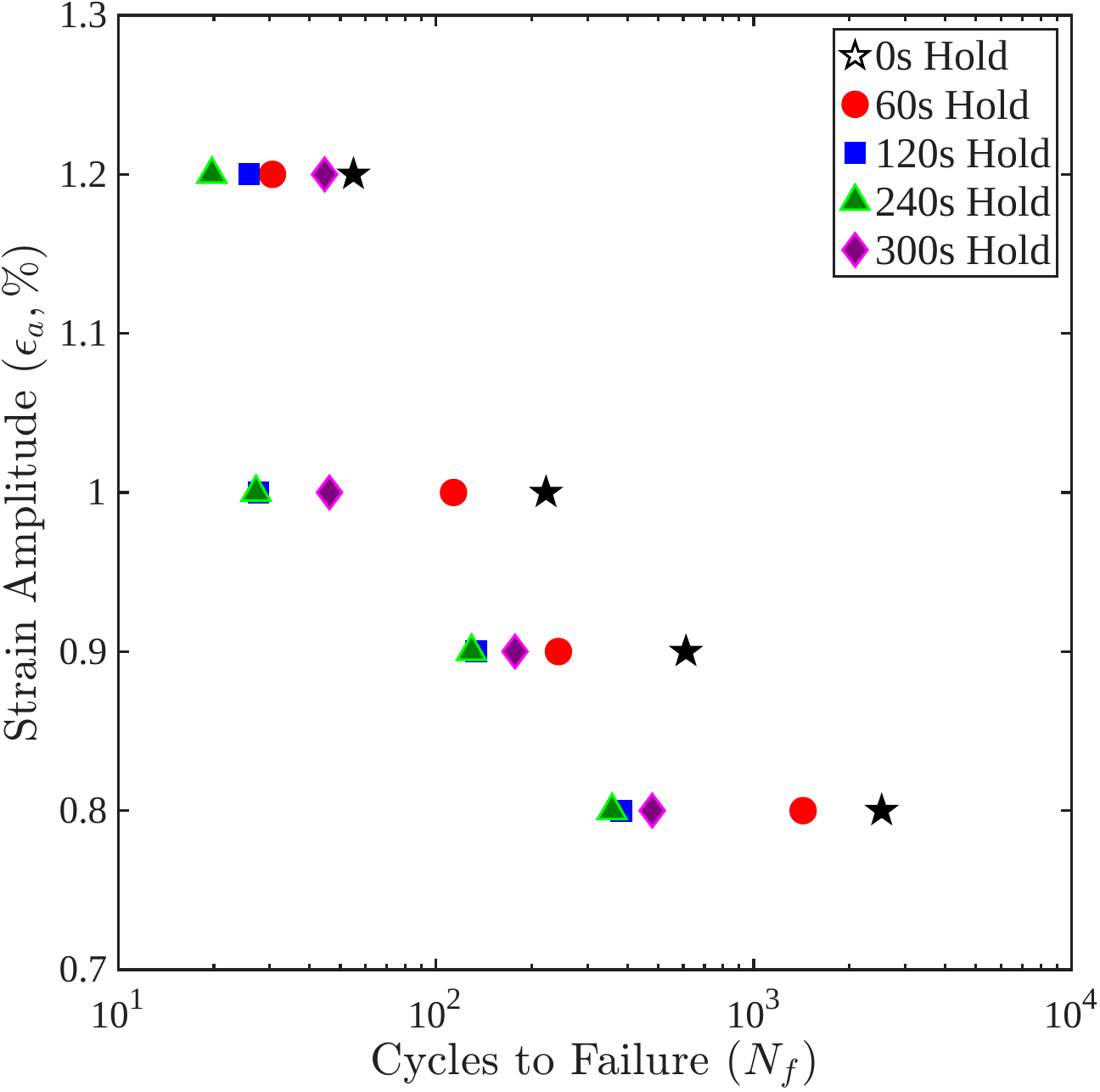}
    \caption{} 
    \label{fig:strain_hold_compare}
\end{subfigure}

    \vspace{0.4cm} 
    \captionsetup{justification=justified} 
    \caption{Simulation of the cyclic deformation behavior.(a) Creep-fatigue stress-strain response at a strain amplitude of 1.2\% under various tensile holding (TH) conditions from $0\,s$ to $300\,s$. The stress relaxation profiles of the $10^{th}$ cycle are presented in (b), showing the stress decay for different hold periods, while (c) illustrates the resulting relationship between the predicted number of cycles to failure ($N_{f}$) and the tensile holding time across three distinct strain amplitudes and (d) Fatigue life at various holding periods and strain amplitudes.}
    \label{fig:Ghen_model_1}
\end{figure}

The effect of hold time and strain-amplitude on the cyclic stress-strain response and predicted life is shown in \cref{fig:Ghen_model_1}. The creep-fatigue interaction response is shown with the stabilized hysteresis loops at the $10^{th}$ cycle in \cref{fig:holdtime}, at a constant strain amplitude of 1.2\% over a range of tensile hold (TH) durations from $0~s$ (corresponding to pure-fatigue) to $300~s$. The findings demonstrate a notable stress relaxation during the dwell period, which increases with increase in hold time and results in a increase of  inelastic creep strain accumulation.

\cref{fig:life_vs_mean_2} shows stress relaxation profiles for the $10^{th}$ cycle. It can be seen from these plots that during the tensile hold, stress drops quickly with hold time at first and then reduces gradually with further hold. Although the stress-relaxation curve for hold time of $30\,s$ is distinct, the stress relaxation profiles for hold times ranging from $60\,s$ to $300\,s$ are very close to each other. This shows that the early part of the hold period has the maximum effect on the stress relaxation and holds beyond $60\,s$ has minimal change to the response. The observed stress relaxation during the tensile hold period results from the conversion of elastic strain into inelastic creep strain during this stage. This time-dependent strain accumulation promotes microstructural degradation through the nucleation and growth of intergranular cavities, significantly accelerating the damage compared to pure fatigue conditions which is discussed in \cite{DING2018319}. \cref{fig:life_vs_ratio_2} shows the link between creep-fatigue life and the tensile holding duration across three different strain amplitudes ranging from 0.8\% to 1.0\%. It is evident from the bar chart that increasing the strain amplitude at any given hold duration makes the number of cycles to failure $N_{f}$ reduce significantly. For example, during a hold duration of $60\,s$, $N_f$ reduces from  2270 cycles at $\epsilon_a=0.8\%$ to only 100 cycles at $\epsilon_a= 1.0\%$. Interestingly, at a given strain amplitude, the life initially decreases rapidly with an increase in hold time (till around $60\,s$) followed by a more gradual decrease till around $240\,s$ when it reaches the minimum value. Further increase in hold time leads to a slight increase in predicted life. Such an effect of hold time on the creep-fatigue life has been observed in the experimental study of \cite{CHEN2016175}.

    

\begin{figure}[h!]
    \centering
     \captionsetup{justification=justified}
    \includegraphics[width=0.43\textwidth]{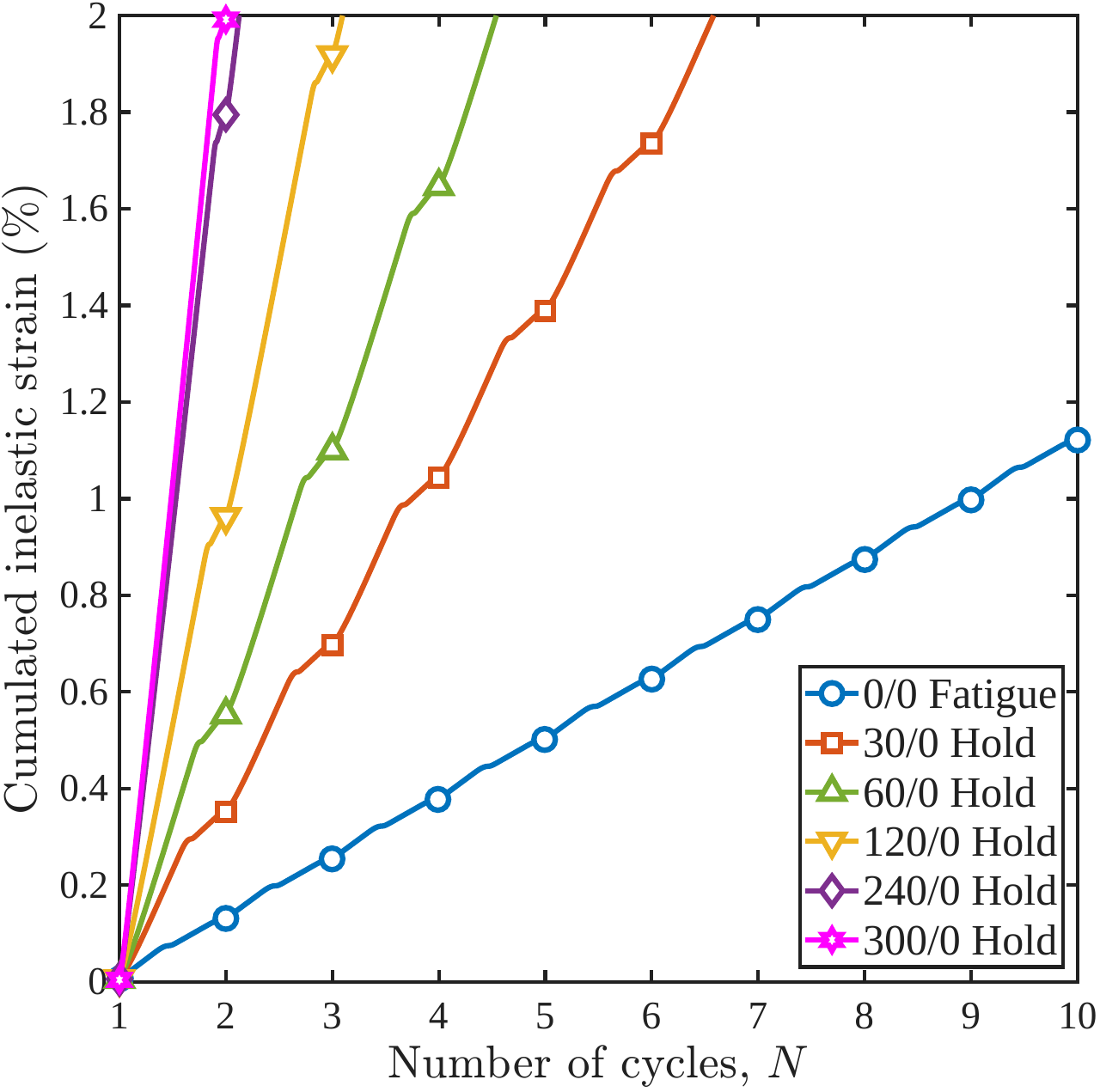}
    \caption{Evolution of cumulated inelastic strain at 760 °C. The plot illustrates the accelerated damage accumulation as tensile dwell time increases from $0\,s$ to $300\,s$, highlighting the transition from pure fatigue to dominant creep-fatigue interaction (CFI) over the first 10 cycles.}
    \label{fig:cumulative_strain}
\end{figure}

In \cref{fig:strain_hold_compare}, the effect of both tensile hold time and the strain amplitude is discussed. The fatigue life decreases sharply with the increase
of strain amplitude in LCF tests. Compared to the fatigue life at the same strain amplitude in LCF tests, the creep-fatigue life decreases when tensile holding periods are introduced. An increase in strain amplitude and tensile hold duration collaboratively decreases $N_{f}$, as observed by the leftward shift of data points and a similar trend is also observed in \cite{CHEN2016175}.

The evolution of accumulated inelastic strain under different hold times is shown in Fig. \ref{fig:cumulative_strain}. To study the effect of hold time on the accumulated inelastic strain, the simulations were continued until the hysteresis loop reached saturation at the $10^{th}$ cycle. As seen in the figure, when the hold time increases, the accumulated inelastic strain also increases and leads to more damage. It is observed that damage is more pronounced between $30\,s$ to $240\,s$, as the initial dislocation structure is unstable and prone to further deformation. In contrast, from  $240\,s$ to $300\,s$, the difference in damage is mild as the dislocation network reaches a more stable configuration, a phenomenon also discussed by \cite{WANG2020105879}.
 


\subsubsection{Interaction of creep and fatigue damage}

\begin{figure}[h!] 
    \centering
    \captionsetup[subfigure]{format=plain, justification=centering, singlelinecheck=false}
    
    \begin{subfigure}[b]{0.47\textwidth}
        \centering
        \includegraphics[width=\textwidth]{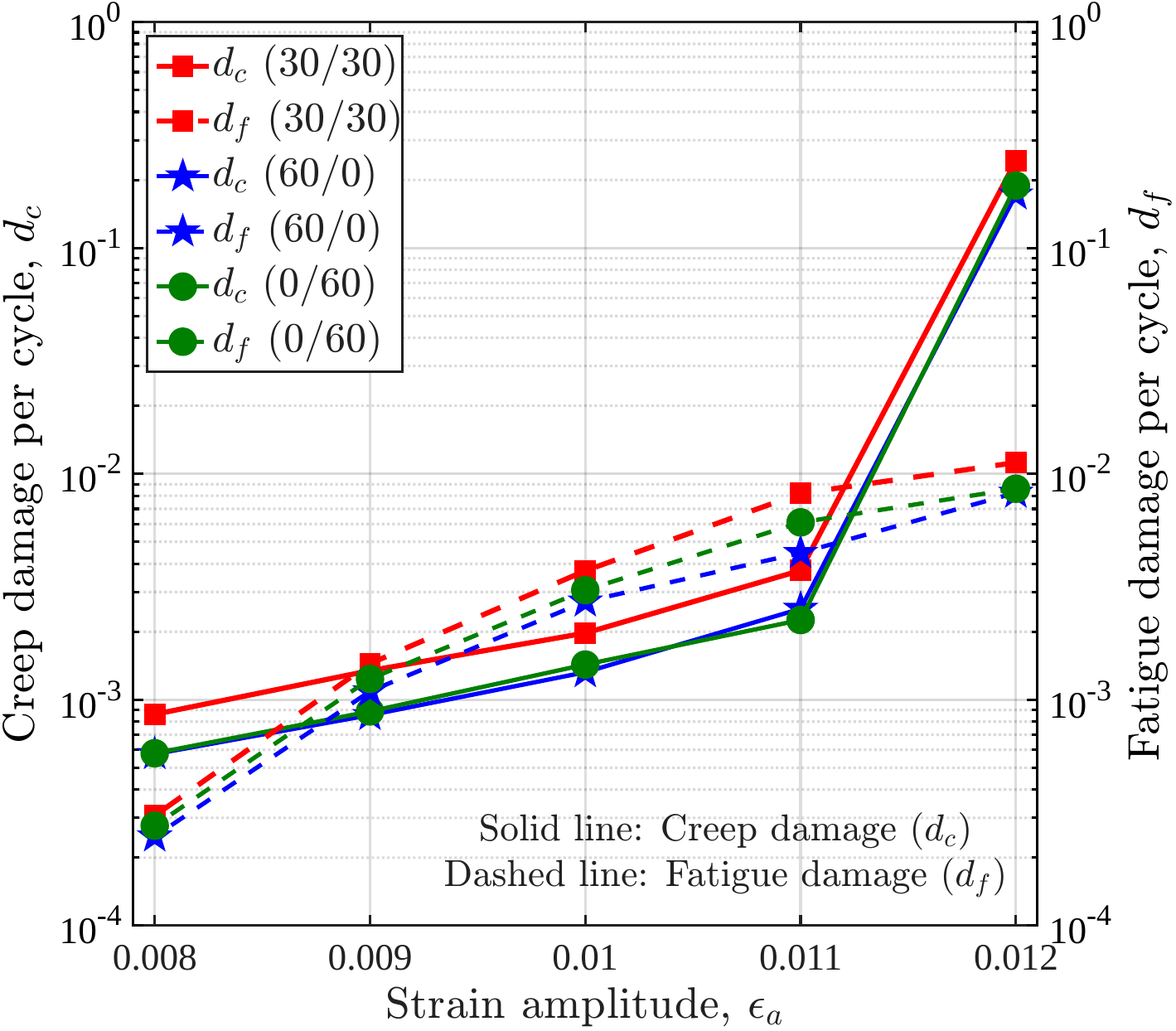}
        \caption{} 
        \label{fig:fatigue_R_new}
    \end{subfigure}
    \hfill
    \begin{subfigure}[b]{0.488\textwidth}
        \centering
        \includegraphics[width=\textwidth]{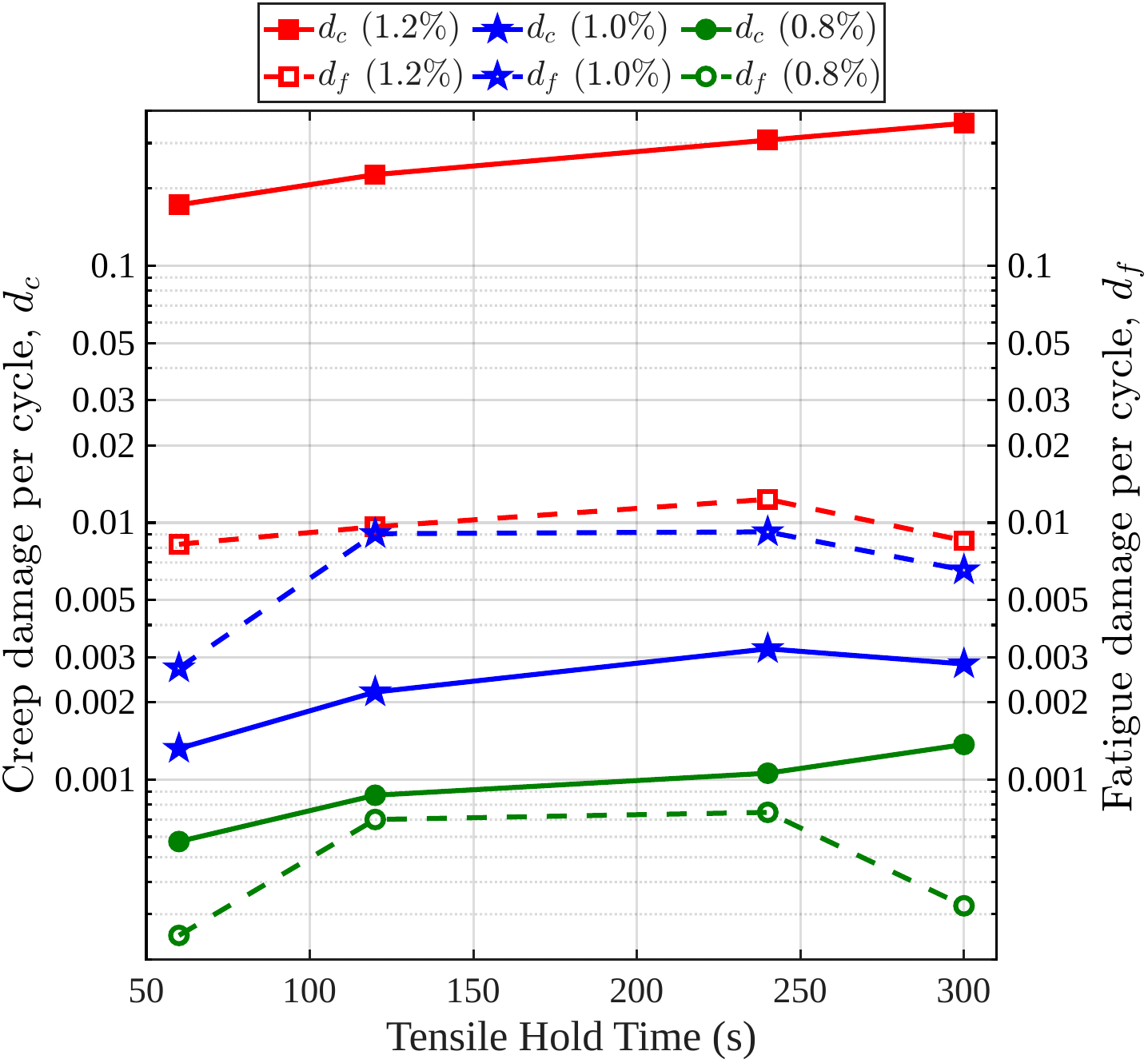} 
        \caption{}
        \label{fig:damage_hold_time_new}
    \end{subfigure}
 
    \vspace{0.4cm} 
    \captionsetup{justification=justified} 
    \caption{Analysis of damage components: (a) Influence of strain amplitude at ${30/30}\,s$, ${60/0}\,s$ and ${0/60}\,s$ hold on creep and fatigue damage accumulation; (b) Evolution of damage components with tensile hold duration at 1.2\%, 1.0\% and 0.8\% strain amplitude.}
    \label{fig:damage_analysis_creep_fatigue}
\end{figure}

The effects of strain amplitude and hold time on the evolution of fatigue and creep damage are discussed next. In \cref{fig:fatigue_R_new}, both damage components have a non-linear dependence on the strain amplitude which is calculated at a hold of $30/30\,s$. At lower strain amplitude ($\epsilon_{a}=0.7\%)$, time-dependent creep damage governs the cyclic degradation, while fatigue damage remains relatively minimal. As the strain amplitude increases towards 0.9\%, the rate of fatigue damage accumulation accelerates rapidly, resulting in state where both fatigue and creep mechanisms contribute equally to the per-cycle damage. Interestingly, at the highest strain amplitude $(\epsilon_{a}=1.2\%)$ considered in this work, the creep damage component increases rapidly and exceeds fatigue damage by over an order of magnitude.

\cref{fig:damage_hold_time_new} shows the effect of tensile hold duration on the individual damage components. Under pure fatigue conditions ($0\,s$ hold time), only baseline fatigue damage is present. However, the introduction of a hold period not only activates creep damage but also increases the fatigue damage per cycle. As the hold time increases from $60\,s$ to $240\,s$, both $d_{c}$ and $d_{f}$ rise. It is documented in literature (\cite{WANG2020105879}) that creep voids and plastic
dimples gradually increases with hold time. Therefore, with increasing hold duration, creep and fatigue damage mechanisms happen collaboratively and creep voids promote fatigue crack growth. Beyond $240\,s$ tensile hold, a saturation effect is observed and both damage rates show a slight decrease at $300\,s$.

\begin{figure}[h!]
    \centering
     \captionsetup{justification=justified}
    \includegraphics[width=0.6\textwidth]{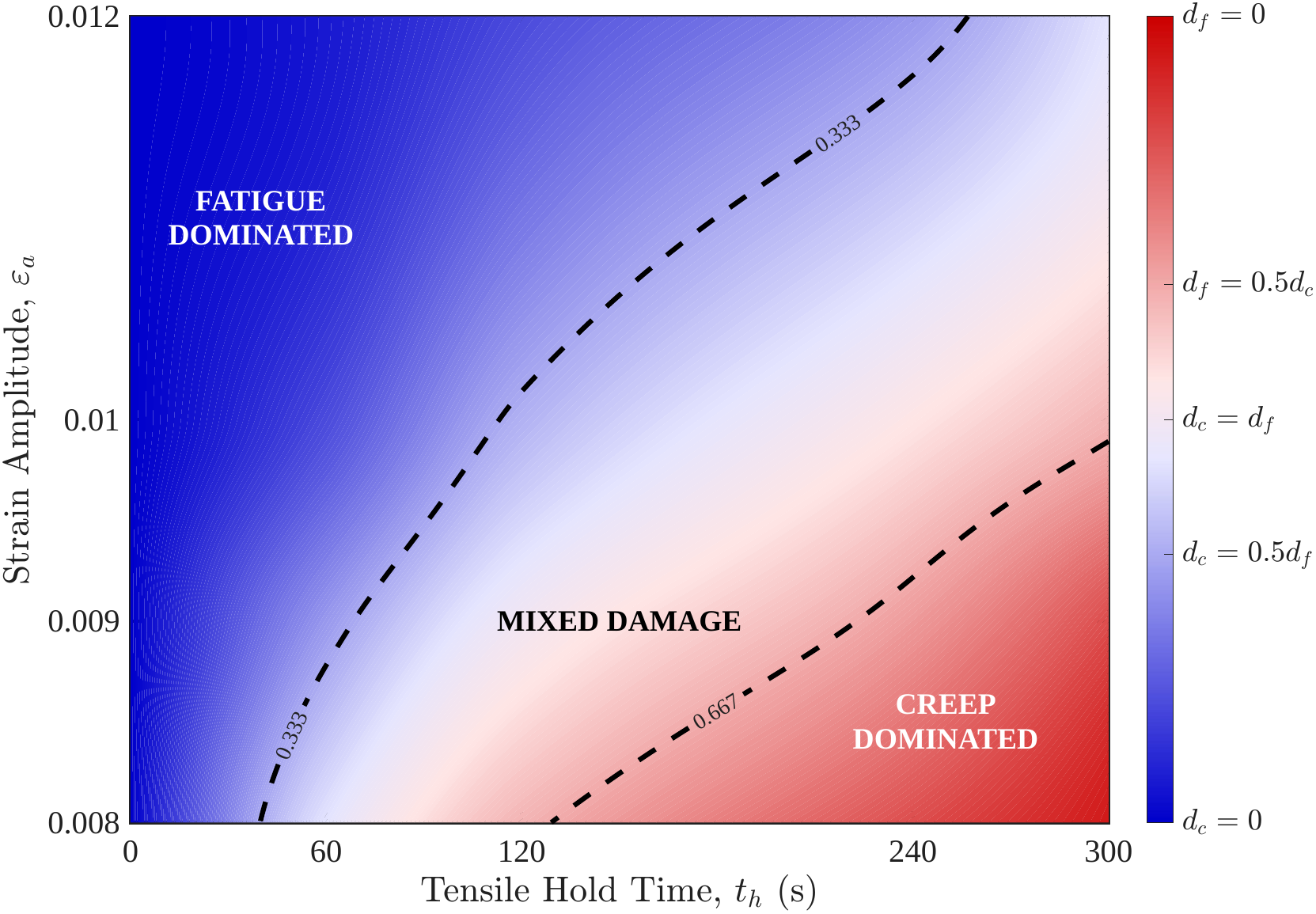}
    \caption{Damage mechanism map at 760° C, showing regimes of fatigue-dominated, creep-dominated, and mixed-mode damage as a function of strain amplitude and hold time. }
    \label{fig:damage_map_strain_amp_hold}
\end{figure}

The damage mechanism map in \cref{fig:damage_map_strain_amp_hold} shows a transition in the failure mode as a function of tensile hold time ($t_h$) and strain amplitude ($\epsilon_a$). For the partitioning of the strain-amplitude and hold time space into fatigue and creep-dominated regions, a damage-ratio parameter (Z = $\frac{d_c}{d_c +d_f}$) is used. The region where $Z < 0.333$ (or equivalently $d_f > d_c$) is fatigue dominated zone, where cyclic plasticity is the main reason for degradation. The mixed damage region ($0.333\le Z \le 0.667$) shows a competitive interaction between cyclic and time dependent mechanisms. For $Z \geq 0.667$ ($d_f < d_c$), the damage mode is creep dominated and time dependent. We can conclude based on this map that at low hold times, the damage is fatigue dominated across all strain amplitudes, while at long hold times and small to intermediate strain amplitudes, the damage is creep-dominated. The transition between these two regimes is characterized by comparable contributions from creep and fatigue.

\subsection{Mean strain and R-ratio effects}

\begin{figure}[h!]
    \centering
    \captionsetup[subfigure]{format=plain, justification=centering, singlelinecheck=false}
    
    \begin{subfigure}[b]{0.44\textwidth}
        \centering
        \includegraphics[width=\textwidth]{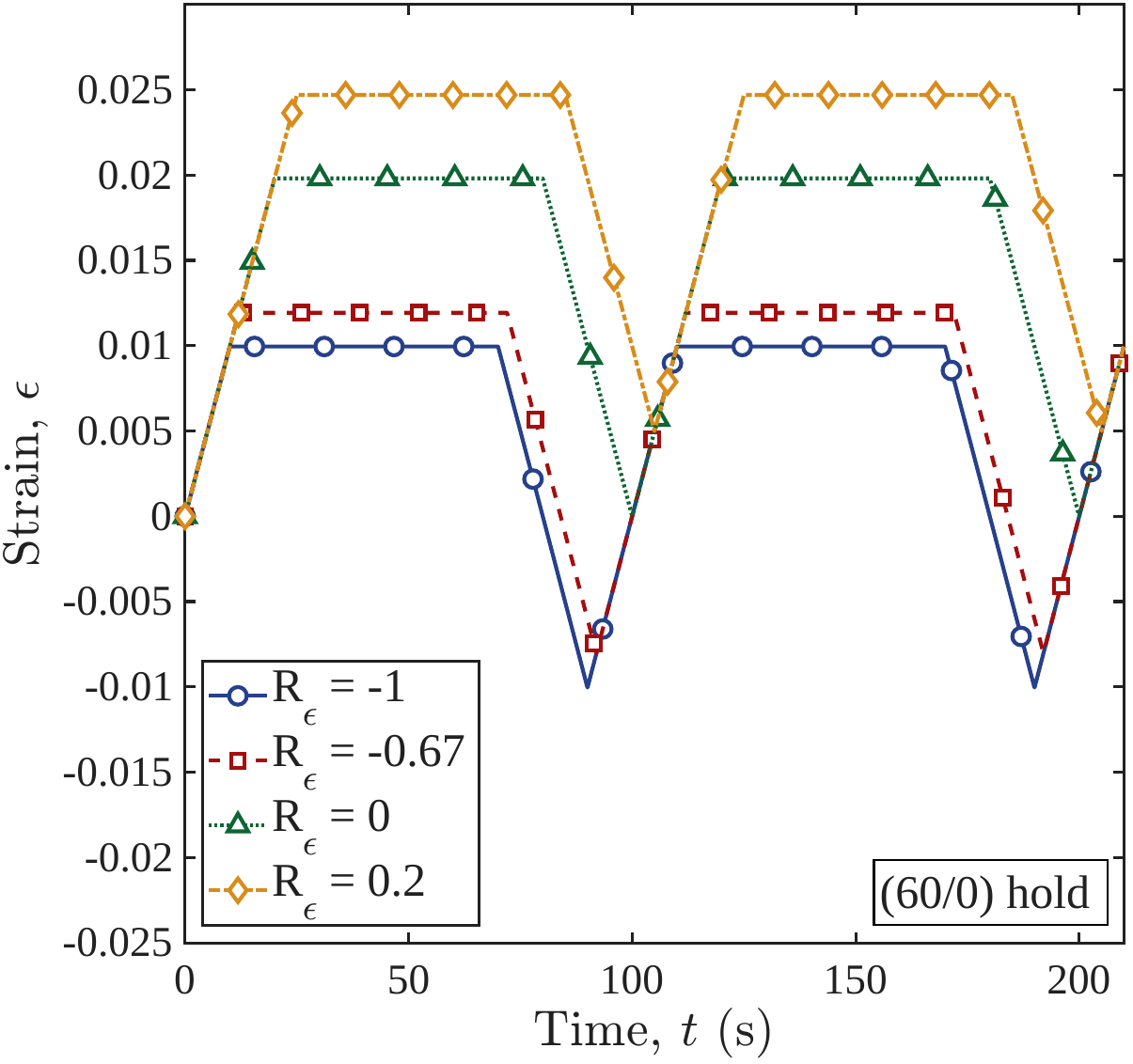}
        \caption{} 
        \label{fig:loading_waveform_11}
    \end{subfigure}
    \hfill
    \begin{subfigure}[b]{0.45\textwidth}
        \centering
        \includegraphics[width=\textwidth]{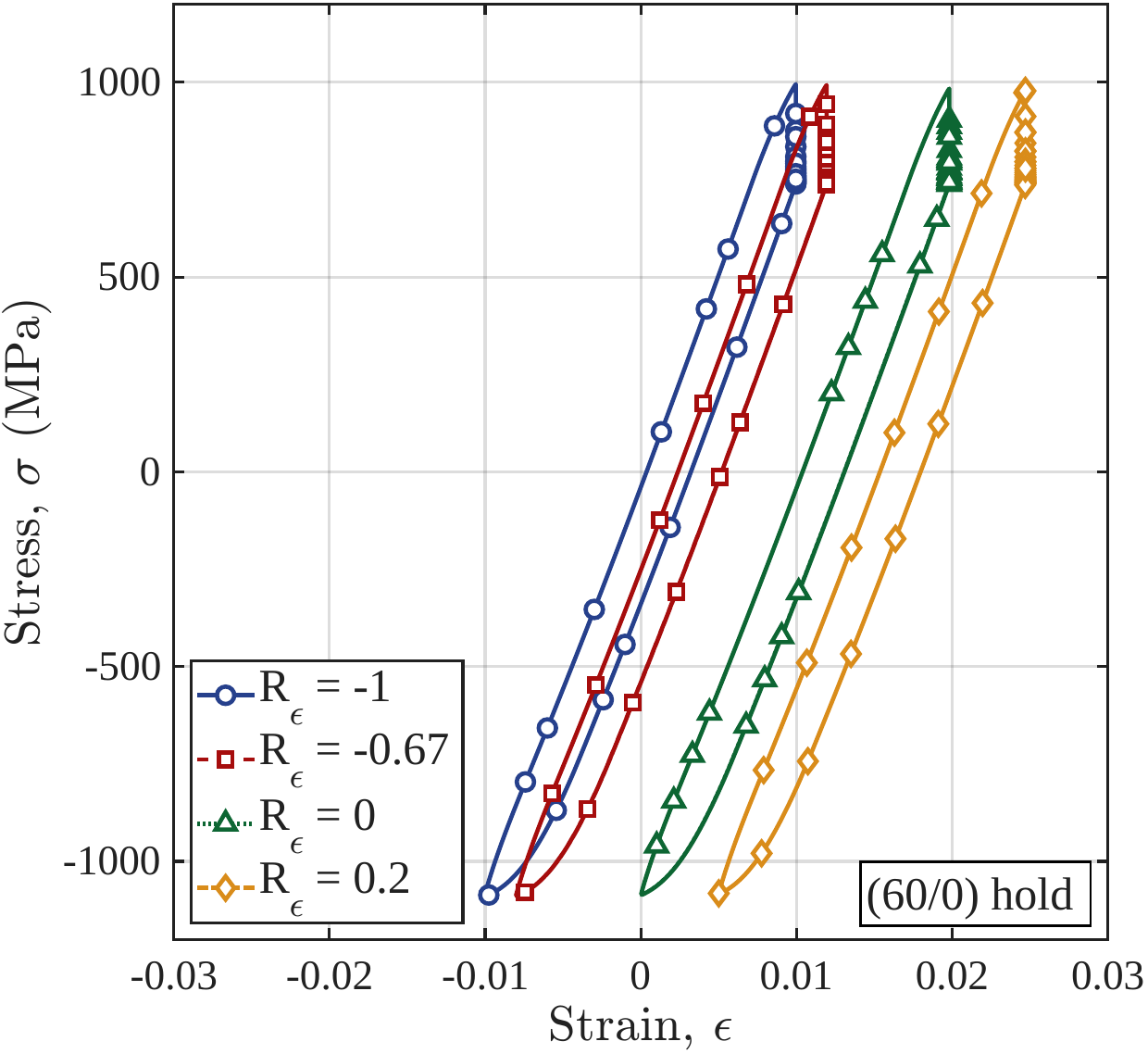}
        \caption{} 
        \label{fig:hysteresis_loops_11}
    \end{subfigure}
     \caption{Effect of strain ratio ($R_{\epsilon}$) on (a) loading waveforms and (b) stabilized hysteresis loops. All cases include a $60\,s$ tensile hold period .}
    \label{fig:R_ratio_Effect}
\end{figure}

\cref{fig:R_ratio_Effect} shows the variation of predicted life with strain ratio $R_{\epsilon}=\frac{\epsilon_{min}}{\epsilon_{max}}$ at different hold configurations. The loading waveform is shown in \cref{fig:loading_waveform_11}, and the stabilized stress-strain hysteresis loop for the $10^{th}$ cycle are presented in \cref{fig:hysteresis_loops_11}. As $R_\epsilon$ increases from -1 (fully reversed) to 0.2 (tension-biased), the stabilized hysteresis loop shifts to the right as shown in \cref{fig:hysteresis_loops_11} and the mean strain becomes tensile. 

\begin{figure}[h!]
    \centering
    \captionsetup[subfigure]{justification=centering}
    
    \begin{subfigure}[t]{0.424\textwidth}
        \centering
        \includegraphics[width=\textwidth]{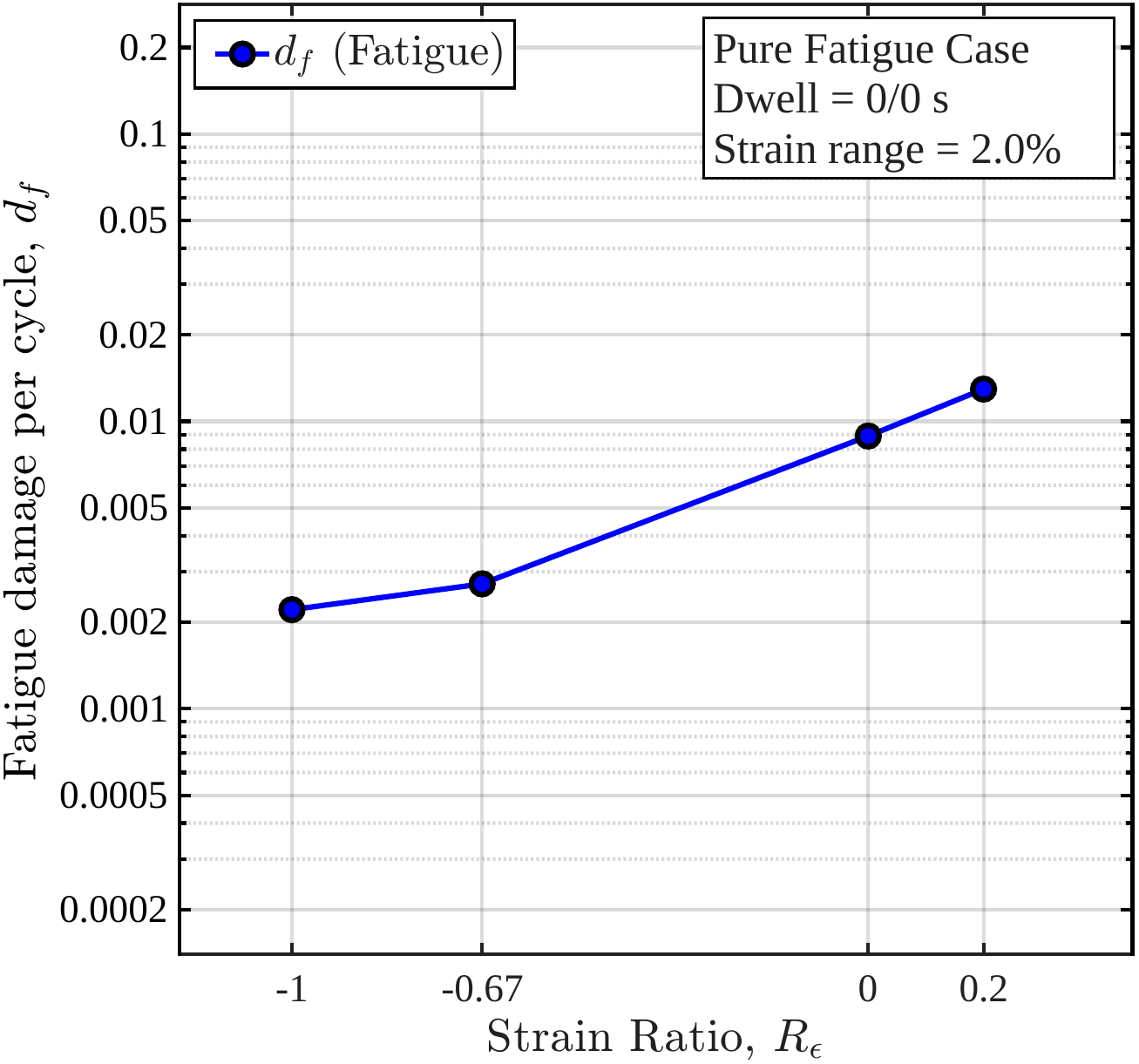}
        \caption{} 
        \label{fig:fatigue_R}
    \end{subfigure}
    \hfill 
    \begin{subfigure}[t]{0.5\textwidth}
        \centering
        \includegraphics[width=\textwidth]{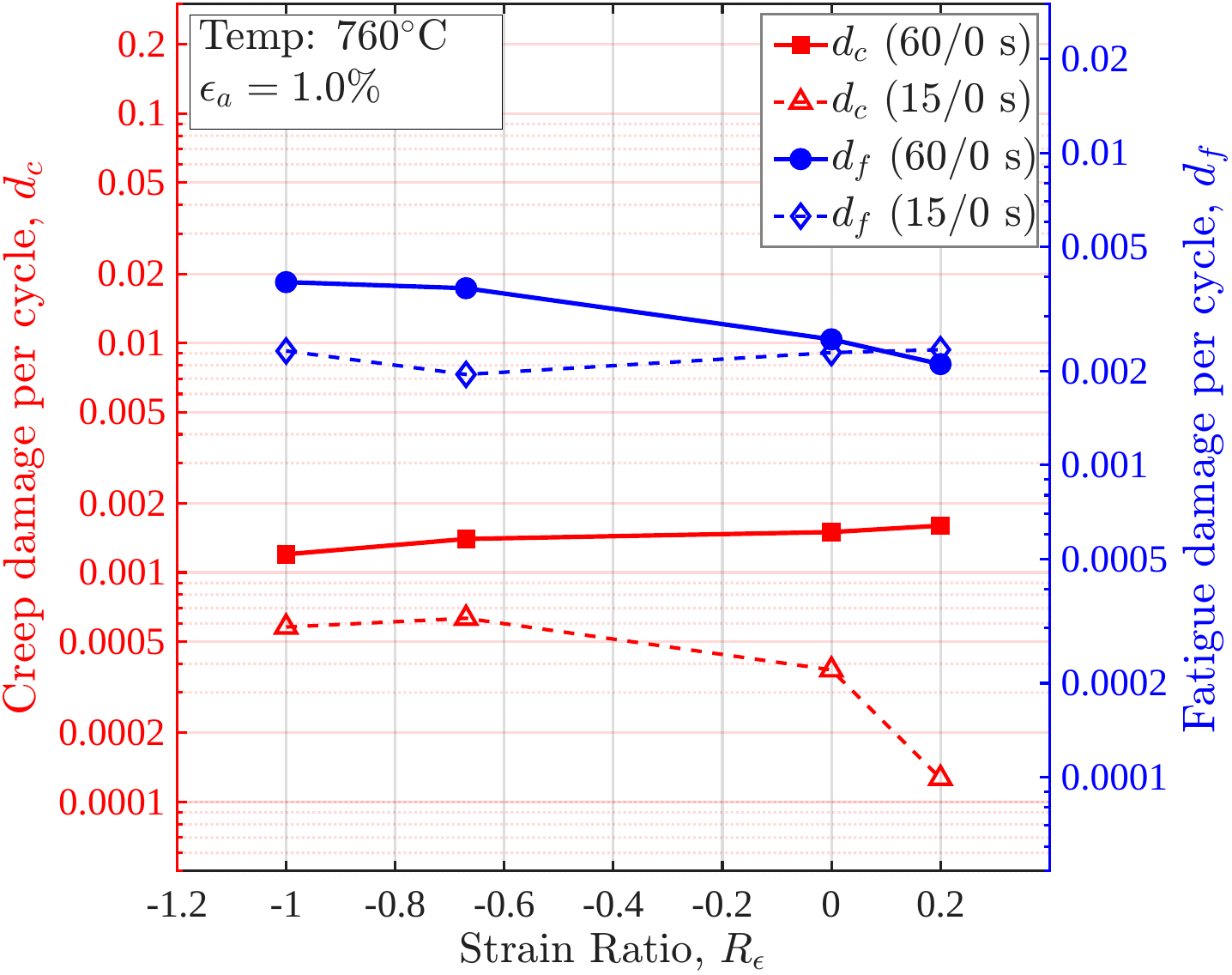}
        \caption{} 
        \label{fig:creep_R}
    \end{subfigure}
    
    \vspace{0.4cm} 
    
    \begin{subfigure}[t]{0.435\textwidth}
        \centering
        \includegraphics[width=\textwidth]{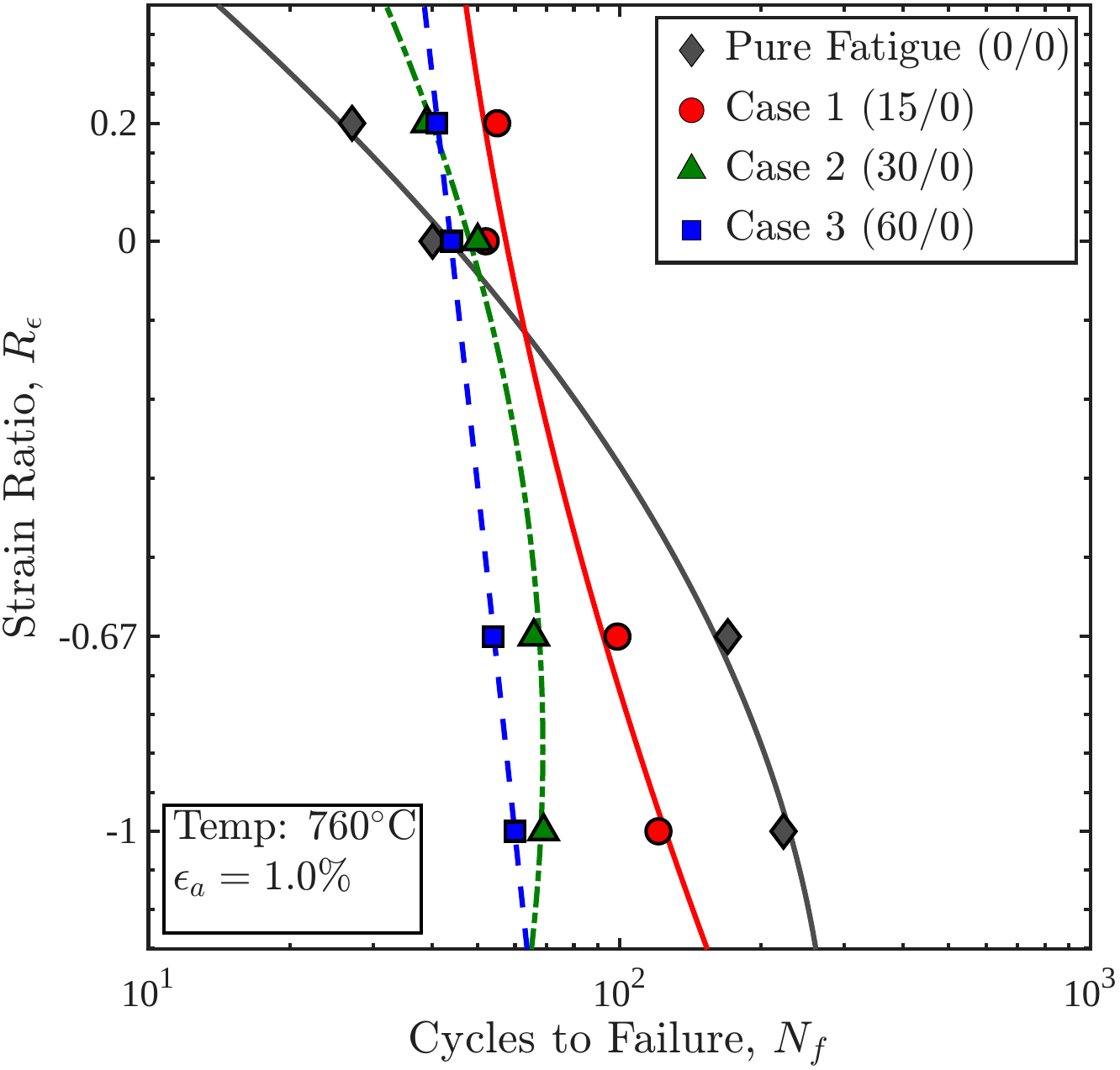}
        \caption{} 
        \label{fig:R_effect}
    \end{subfigure}

    \vspace{0.4cm} 
    
    \caption{Comprehensive numerical simulation results under varying loading conditions: (a) Effect of $R_{\epsilon}$ ratio on accumulated pure fatigue damage ($d_f$). (b) Creep damage ($d_c$) and fatigue damage ($d_f$) per cycle for $15/0\, s$ and $60/0\, s$ hold cases; (c) Predicted cyclic life ($N_f$) as a function of strain ratio ($R_{\epsilon}$).}
    \label{fig:mean_strain_analysis_1}
\end{figure}

In \cref{fig:fatigue_R}, we observe that as $R_\epsilon$ increases, the fatigue damage per cycle increases rapidly in pure fatigue at 1.0\% strain amplitude due to increasing tensile bias in the loading waveform, which accelerates crack nucleation and propagation.  
On the other hand, adding a hold time not only causes stress relaxation and creep damage but also increases fatigue damage, as evident in \cref{fig:creep_R}, demonstrating the strong interaction between the two damage mechanisms. In pure fatigue, when $R_\epsilon$ increases from -1 to 0.2, the fatigue life reduces, as higher tensile mean strain accelerates damage. But the introduction of $15\,s$ tensile hold leads to significant reduction in creep-fatigue life, as seen in \cref{fig:R_effect}. With further increase of tensile hold, the reduction in life with $R_\epsilon$ is comparatively milder compared to that in pure fatigue. A similar observation has been made in the experimental results of \cite{SUN2021106187}.

\subsection{Effect of temperature}

\begin{figure}[h!]
    \centering
     \captionsetup[subfigure]{format=plain, justification=centering, singlelinecheck=false}
    \begin{subfigure}[b]{0.425\textwidth}
        \centering
        \includegraphics[width=\textwidth]{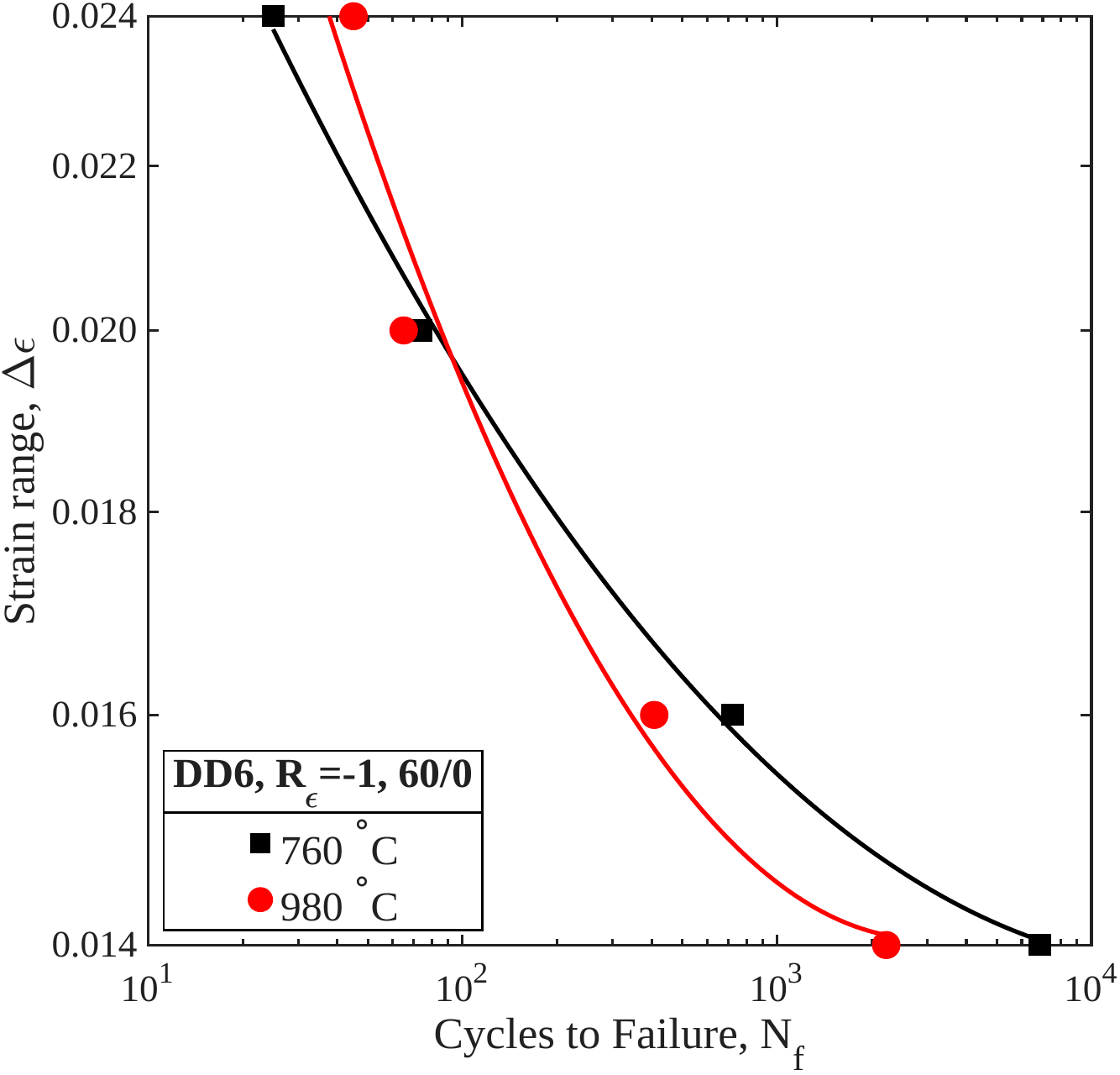}
        \caption{}
        \label{fig:life_980}
    \end{subfigure}
    \hfill 
    \begin{subfigure}[b]{0.425\textwidth}
        \centering
        \includegraphics[width=\textwidth]{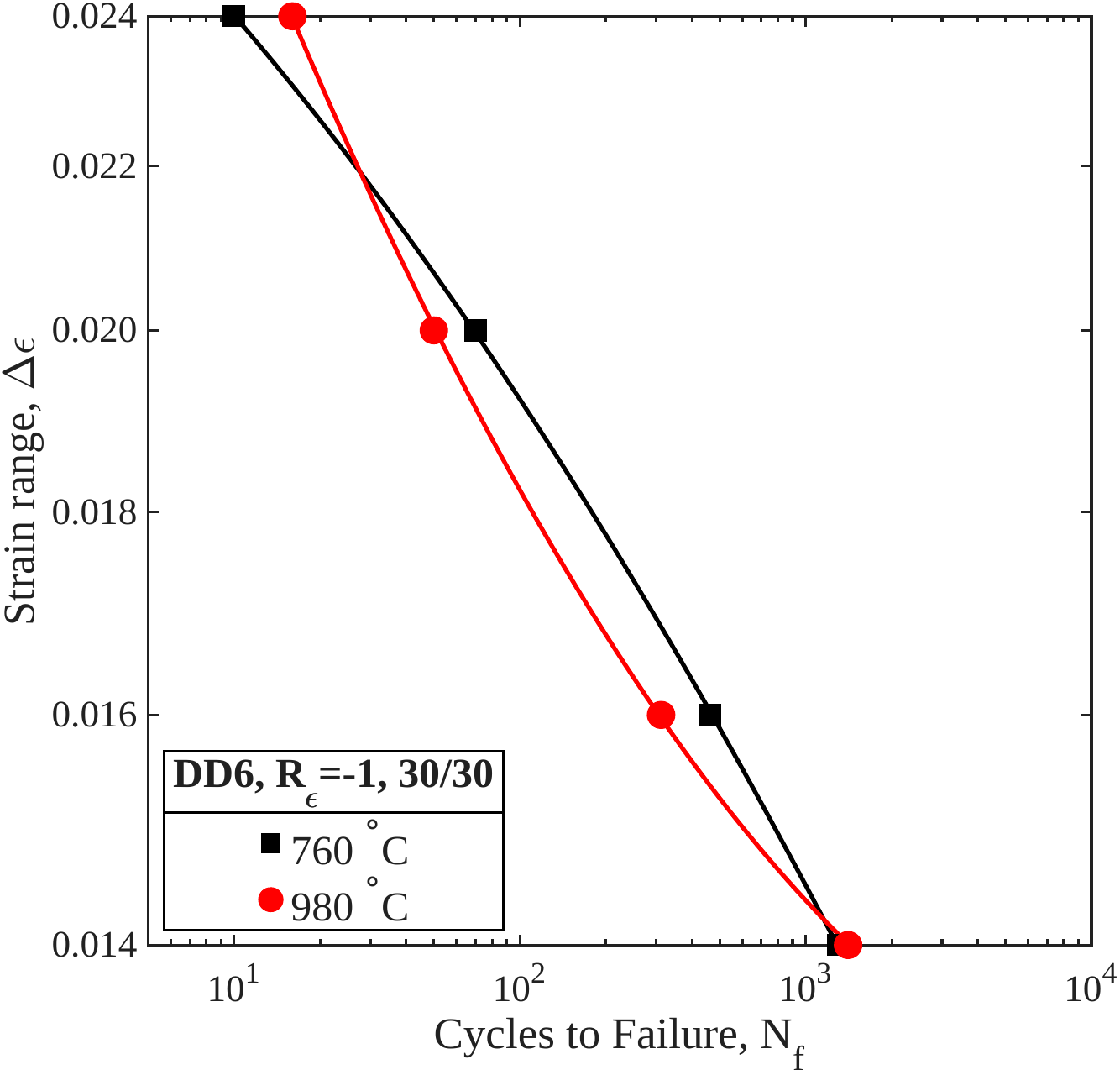}
        \caption{}
        \label{fig:life_980_cons}
    \end{subfigure}
    
    \vspace{0.4cm}
    \captionsetup{justification=justified}
    \caption{Comparison of model predicted creep fatigue life versus strain range at different temperatures (760$^\circ$C and 980$^\circ$C) for hold configurations: (a) 60/0 and (b) 30/30.}
    \label{fig:760_980_plot}
\end{figure}

The temperature dependence of creep-fatigue life is shown in \cref{fig:760_980_plot}. The creep-fatigue life at 760$^{\circ}$C is more than that at 980$^{\circ}$C for most strain ranges except at very high strain ranges ($\geq 2\%$ for 60/0 s hold and $\geq 2.2\%$ for 30/30 s hold respectively). At higher temperature, the rate of dislocation climb, vacancy creation and diffusional creep are high (\cite{Reed2008_CMSX4, ZHU20124888}), which leads to more creep strain and reduction in life. 

\begin{figure}[h!]
    \centering
     \captionsetup[subfigure]{format=plain, justification=centering, singlelinecheck=false}
    \begin{subfigure}[b]{0.42\textwidth}
        \centering
        \includegraphics[width=\textwidth]{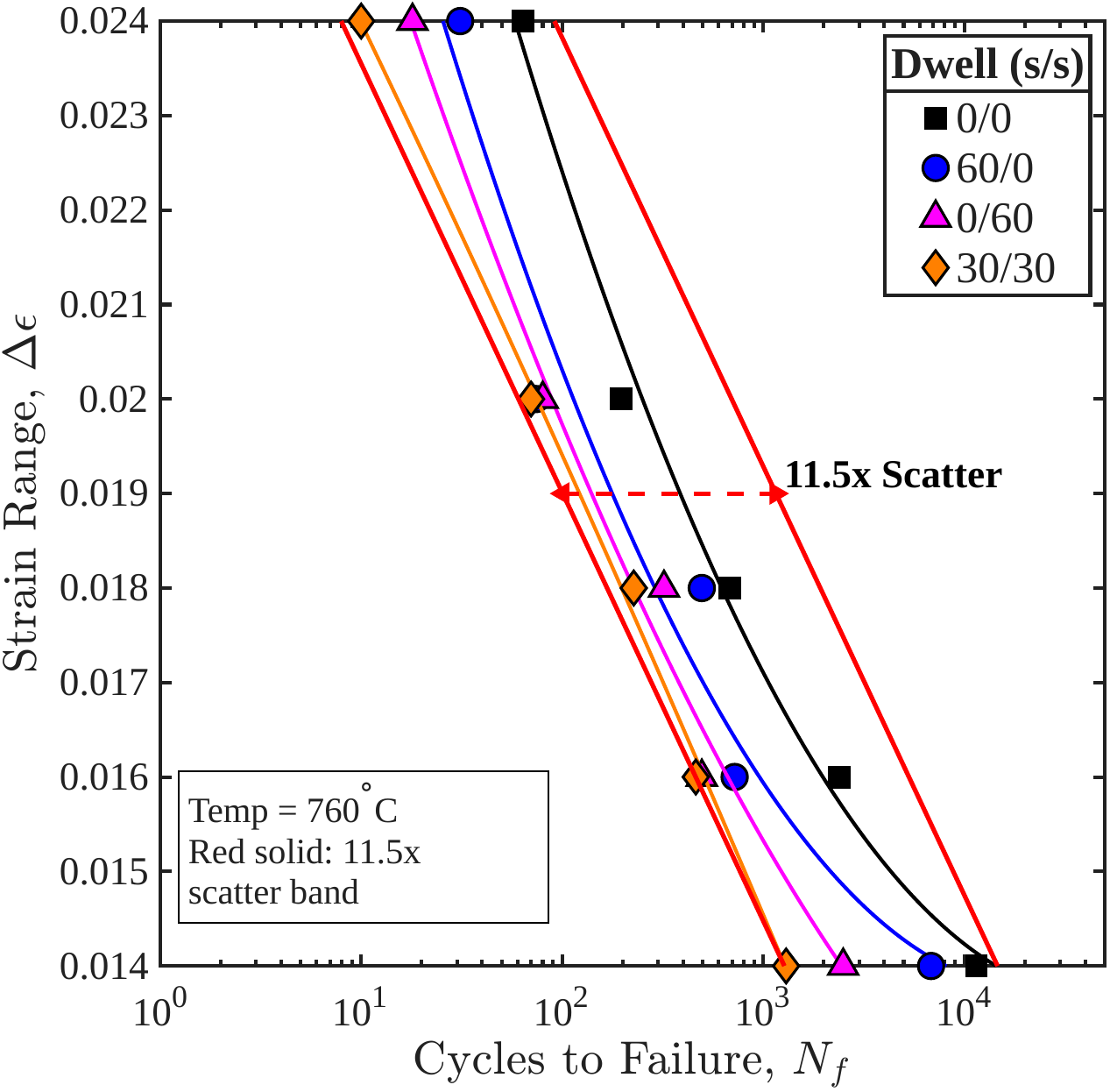}
        \caption{}
        \label{fig:life_980}
    \end{subfigure}
    \hfill 
    \begin{subfigure}[b]{0.42\textwidth}
        \centering
        \includegraphics[width=\textwidth]{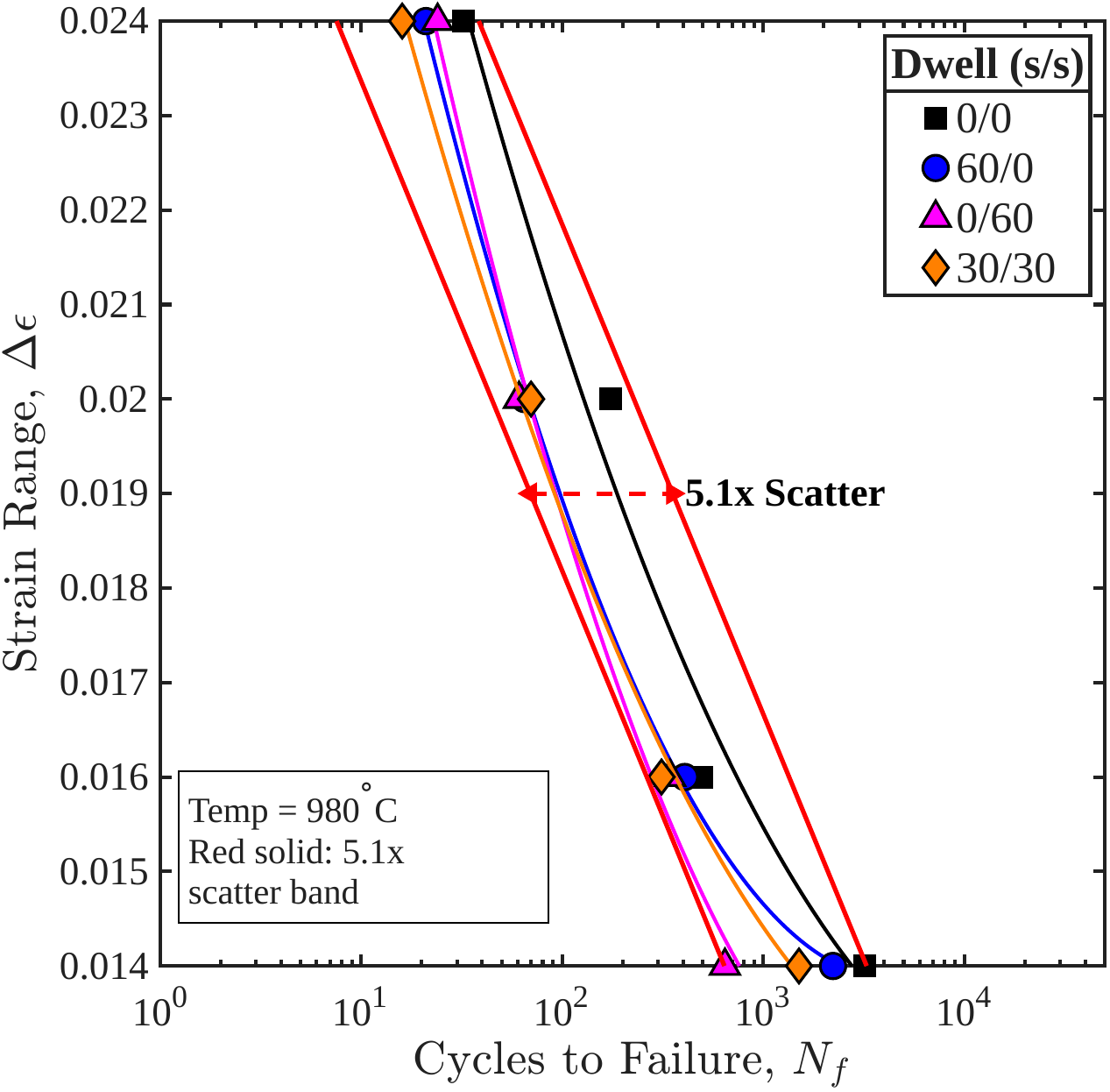}
        \caption{}
        \label{fig:life_980_cons}
    \end{subfigure}
    
    \vspace{0.2cm}
    \captionsetup{justification=justified}
    \caption{Model predicted creep-fatigue life versus strain range for different hold configurations as marked in the legend, at (a) 760$^{\circ}$C and (b) 980$^{\circ}$C. At high temperature, the curves cluster together which leads to a reduction in the scatter band.}
    \label{fig:combined_life_plots}
\end{figure}

\begin{figure}[h!]
    \centering
     \captionsetup{justification=justified}
    \includegraphics[width=0.45\textwidth]{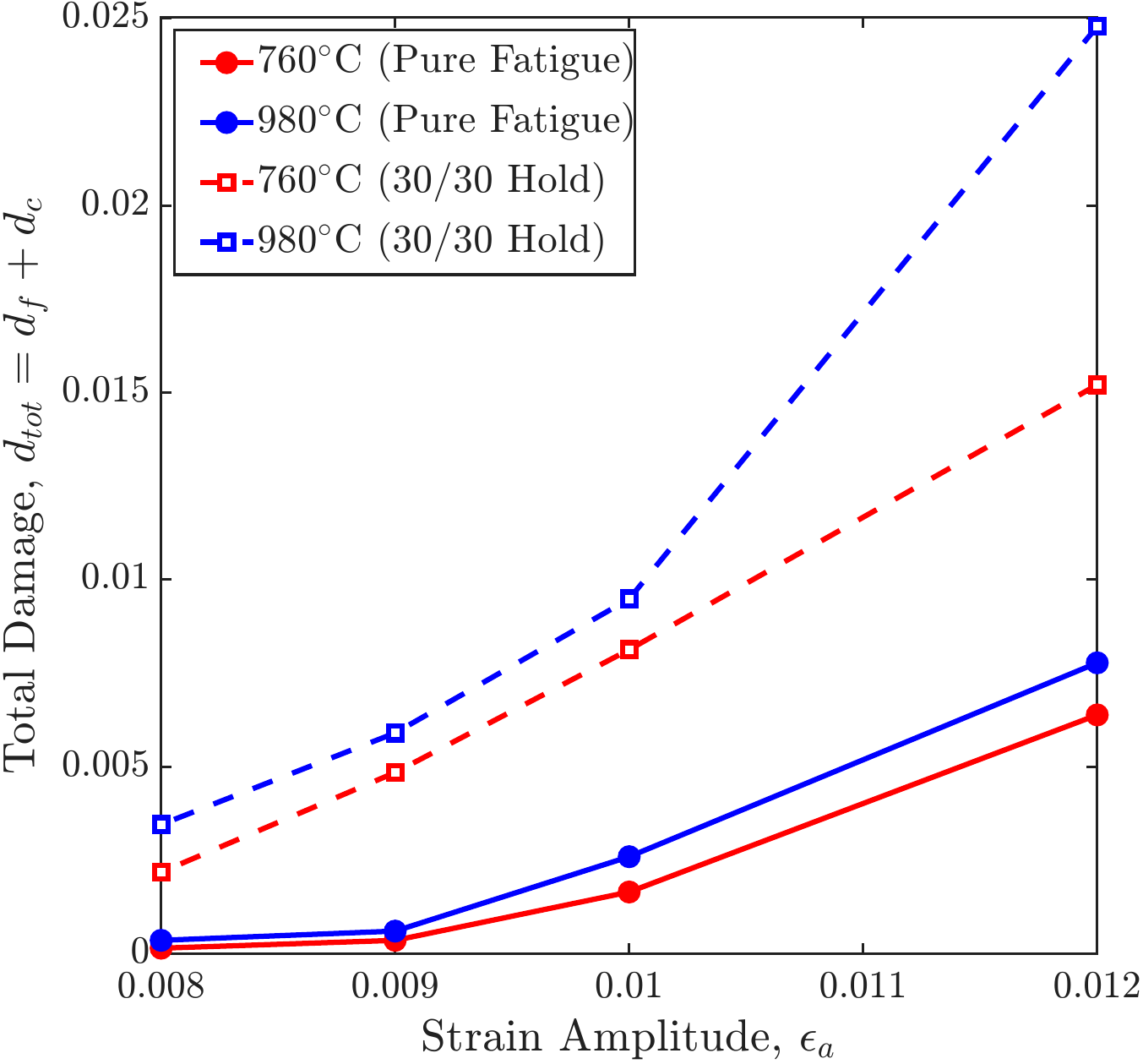}
    \caption{Total damage with pure fatigue  and (30/30) hold configuration at different strain amplitude at (a) 760$^{\circ}$C and (b) 980$^{\circ}$C. }
    \label{fig:temperature_depn_damage }
\end{figure}

The creep-fatigue life as a function of strain amplitude for different hold times at 760$^{\circ}$C and 980$^{\circ}$C are shown in \cref{fig:combined_life_plots}. The life versus strain range plots across all hold types show a scatter band of 5.1 at 980$^{\circ}$C and 11.5 at 760$^{\circ}$C. It has been documented in literature (\cite{Shi2013}) that at elevated temperature, these individual curves begin to cluster much more tightly, which is in agreement with reduction of the scatter band obtained by the model prediction at high temperature.

Compared with pure fatigue, all of the three hold types lead to obvious life degradation. Roughly, balanced hold type shows the shortest life at both 760$^{\circ}$C and 980$^{\circ}$C. In balanced hold configuration, the area of the hysteresis loop is significantly higher as compared to the unbalanced holds (as shown in \cref{fig:three_loops}), which causes highest plastic dissipation and results in the shortest life. The tensile hold promotes creep void nucleation and growth, which further assists fatigue crack initiation. On the other hand, the compressive hold configuration leads to a mean tensile stress, which promotes the nucleation of creep voids and accelerates fatigue crack growth during the tensile loading stage. Consequently, the predicted life in tensile (60/0 s) and compressive (0/60 s) holds are close to each other at both temperatures considered, but significantly higher than that in balanced hold.

\cref{fig:temperature_depn_damage } shows that damage accumulation is more severe at 980$^{\circ}$C compared to 760$^{\circ}$C for both pure fatigue and balanced hold configuration (30/30). In pure fatigue, for 980$^{\circ}$C the yield strength is less, which results in larger inelastic strain and more degradation due to easier dislocation movement. This effect is more pronounced during the 30/30 hold due to the significant stress relaxation at 980$^{\circ}$ C.

\section{Conclusions}
The present work extends the CPFE framework to study the creep-fatigue response of single-crystal nickel superalloys under a broader set of practically relevant thermo-mechanical loading conditions. The model is employed to examine the effects of strain amplitude, R-ratio, and hold duration on cyclic deformation, stress relaxation, and creep-fatigue life. In addition, a temperature-dependent term is introduced into the slip resistance formulation to account for the influence of temperature on deformation and damage evolution. Particular attention is given to the separate contributions of fatigue and creep damage, their interaction during damage accumulation, and the identification of dominant damage regimes through a damage-mechanism map in the strain amplitude-hold time space. The results show that the framework is capable of reproducing experimentally observed trends in hysteresis behavior and creep-fatigue life over a wide range of loading conditions.
The main conclusions are drawn as follows:

\begin{enumerate}
    \item[(1)] The study successfully implements a dislocation density-based crystal plasticity model within the MOOSE finite element framework, coupling it with a thermodynamic damage approach based on entropy generation rate. This model accurately captures the stabilized cyclic stress-strain response, including stress relaxation behaviors, under pure fatigue and various hold-time configurations (e.g., $60/0\,s$, $0/60\,s$, $30/30\,s$) at elevated temperatures.
    
    \item[(2)] The introduction of tensile hold times drastically reduces the number of cycles to failure compared to pure fatigue conditions. Stress relaxation primarily occurs in the early stages of the hold period (up to $60\,s$), converting elastic strain into inelastic creep strain. This accelerated degradation leads to a rapid initial decrease in life, followed by a relatively gradual reduction with further increase of hold time, as the dislocation network stabilizes and recovers.

    \item[(3)] The interplay between creep and fatigue damage is highly dependent on the applied strain amplitude. At lower strain amplitudes (e.g., $\epsilon_a$ = 0.8\%), time-dependent creep damage dictates the failure life. In the range of $\epsilon_a=0.9\%$ to ($\epsilon_a=1.0\%$), creep-fatigue interaction is evident. However, as the strain amplitude is increased further, the creep damage again becomes dominant in the balanced $30/30\,s$ hold configuration.
    
    \item[(4)] Creep-fatigue life is lower at 980°C compared to 760°C. The higher temperature promotes dislocation climb, vacancy creation, and diffusional creep mechanisms, which accelerate creep strain accumulation and lead to earlier failure. Additionally, the life prediction for different hold configurations cluster more tightly at higher temperatures. It shows high sensitivity at low temperature compared to high temperature because at high temperature,
   the creep damage tends to reach saturation, making the material less sensitive to further temperature increase as can be seen from the exponential term $\exp(-Q/RT)$ in the creep damage calculation.

\item[(5)] When comparing predicted life to experimental data, the Linear Damage Summation (LDS) rule provides predictions where the majority of data points fall within acceptable ±2x scatter bands. Conversely, the implemented Non-linear Damage Summation (NDS) rule consistently yields more conservative (shorter) life predictions, providing a safer margin for engineering design applications. 

\end{enumerate}


The implemented microstructure-sensitive crystal plasticity model is suitable for application to study component level damage and life prediction under creep-fatigue loading, given its relatively lightweight nature. At the same time, we have demonstrated strong predictive capability of the model by extensive validation against experimental data at different temperatures and loading parameters. In the future, we plan to include more details in the crystal plasticity model to account for the effect of misfit stresses, cubic slip, non-Schmid effects and the effect of microstructural features such as precipitate size, volume fraction and channel width in a more rigorous manner. Future work will also involve application of the model for microstructure-sensitive prediction of creep-fatigue damage and life of critical components such as turbine blades and discs used in gas turbine engines.

\section*{CRediT authorship contribution statement}

\textbf{Santosh Kumar Shaw:} Methodology, Software, Formal analysis, Investigation, Validation, Visualization, Writing – original draft. \textbf{Sabyasachi Chatterjee:} Supervision, Conceptualization, Methodology, Investigation, Writing – original draft, review \& editing. \textbf{Alankar Alankar:} Methodology, Investigation,  Writing – review \& editing. \textbf{Ayan Bhowmik:} Validation, Writing – review \& editing. 

\section*{Declaration of competing interest}
The authors declare that they have no known competing financial interests or personal relationships that could have appeared to influence the work reported in this paper.

\section*{Acknowledgments}
\textbf{S. Shaw} acknowledges the Institute Fellowship received from Indian Institute of Technology Delhi and the High Performance Computing facility at Indian Institute of Technology Delhi. \textbf{S.  Chatterjee} acknowledges the New Faculty Seed Grant and Equipment Matching Grant received from Indian Institute of Technology Delhi, the financial support received from the Anusandhan National Research Foundation (ANRF, erstwhile SERB) via grant no. SRG/2022/001328 and the financial support received from the Department of Applied Mechanics at Indian Institute of Technology Delhi.

\appendix
\section{Global Equilibrium and Newton-Raphson Iteration}
\label{app:A}

In nonlinear structural mechanics, equilibrium at a given load increment is achieved when the internal force vector $\mathbf F_{int}$ equals the external force vector $\mathbf F_{ext}$. The difference between these forces is expressed through the residual vector $\mathbf{R}(\mathbf{u})$: 
\begin{equation}
    \mathbf{R}(\mathbf{u}) = \mathbf{F}_{int} - \mathbf{F}_{ext}= \mathbf{0}
\end{equation}
\begin{equation}
    \mathbf{R}(\mathbf{u}) =  \int_{\Omega} \mathbf{B}^T \boldsymbol{\sigma}(\mathbf{u}) \, d\Omega - \int_{\Omega} \mathbf{N}^T \mathbf{b} \, d\Omega + \int_{\Gamma} \mathbf{N}^T \mathbf{t} \, d\Gamma = \mathbf{0}
\end{equation}
where $\mathbf{B}$ is the strain-displacement matrix and $\bm{\sigma}$ is the Cauchy stress. The external force vector $\mathbf{F}_{ext}$ is written by considering the contributions from body forces $\mathbf{b}$ and surface tractions $\mathbf{t}$.

Since $\mathbf{F}_{int}$ depends non-linearly on the displacement $\mathbf{u}$, an iterative Newton-Raphson scheme is employed to find the solution.

\subsection{Linearization and the Global Stiffness Matrix}
Assuming equilibrium is not satisfied at the current iteration $k$ (i.e., $\mathbf{R}(\mathbf{u}_k) \neq \mathbf{0}$), we seek an iterative correction $\delta \mathbf{u}^{(k+1)}$ such that $\mathbf{R}(\mathbf{u}_k + \delta \mathbf{u}^{(k+1)}) \approx \mathbf{0}$. Performing a first-order Taylor series expansion about the state $\mathbf{u}_k$:

\begin{equation}
    \mathbf{R}(\mathbf{u}_k + \delta \mathbf{u}^{(k+1)}) \approx \mathbf{R}(\mathbf{u}_k) + \left[ \frac{\partial \mathbf{R}}{\partial \mathbf{u}} \right]_{\mathbf{u}_k} \delta \mathbf{u}^{(k+1)} = \mathbf{0}
\end{equation}

The Jacobian of the residual vector is defined as the **Global Tangent Stiffness Matrix** $\mathbf{K}_T$:

\begin{equation}
    \mathbf{K}_T = - \frac{\partial \mathbf{R}}{\partial \mathbf{u}} = \frac{\partial \mathbf{F}_{int}}{\partial \mathbf{u}} = \int_{\Omega} \mathbf{B}^T \mathbb{C}^{alg} \mathbf{B} \, d\Omega
\end{equation}

where $\mathbb{C}^{alg}$ represents the **consistent tangent modulus** derived from the constitutive update at the integration points.

\subsection{Iterative Update and Global Convergence}
The linear system is solved for the displacement correction at each iteration:

\begin{equation}
    \mathbf{K}_T^{(k)} \delta \mathbf{u}^{(k+1)} = \mathbf{R}(\mathbf{u}^{(k)})
\end{equation}

The global displacement field is updated incrementally:

\begin{equation}
    \mathbf{u}^{(k+1)} = \mathbf{u}^{(k)} + \delta \mathbf{u}^{(k+1)}
\end{equation}

The Newton-Raphson iteration continues until the system satisfies the global convergence criterion. To ensure the system has reached a stable equilibrium state, the solver enforces the following force-based residual criterion: 

\begin{equation}
 \|\mathbf{R}(\mathbf{u}^{(k+1)})\| < \epsilon_R \|\mathbf{F}_{ext}\|  
\end{equation}

where $\epsilon_R$ represents the user-defined tolerances for the nonlinear residual (in our case, utilizing absolute and relative tolerances of $5\times10^{-7}$ and $1\times10^{-6}$, respectively). This ensures that the internal stresses and external loads are balanced.

\section{Numerical Implementation of the Constitutive Model}
\label{app:B}

This appendix details the implicit integration scheme and the derivation of the algorithmic tangent stiffness used in the CPFE framework.

\subsection{Local Newton-Raphson for Material Integration}
In the implicit time integration scheme, the shear strain rate $\dot{\gamma}^{(\alpha)}$ on each slip system $\alpha$ at time $t+\Delta t$ is governed by the viscoplastic flow rule:

\begin{equation}
\dot{\gamma}^{(\alpha)} = \dot{\gamma}_0 \left| \frac{\tau_{t+\Delta t}^{(\alpha)} - \chi_{t+\Delta t}^{(\alpha)}}{g_{t+\Delta t}^{(\alpha)}} \right|^{1/m} \text{sgn} \left( \tau_{t+\Delta t}^{(\alpha)} - \chi_{t+\Delta t}^{(\alpha)} \right)
\label{eq:gamma_dot_app}
\end{equation}

To solve for $\dot{\gamma}^{(\alpha)}$ implicitly, a local residual function $f^{(\alpha)}$ is defined for each slip system:

\begin{equation}
f^{(\alpha)} = \tau_{t+\Delta t}^{(\alpha)} - \chi_{t+\Delta t}^{(\alpha)} - g_{t+\Delta t}^{(\alpha)} \left( \frac{\dot{\gamma}^{(\alpha)}}{\dot{\gamma}_0} \right)^m \text{sgn} \left( \dot{\gamma}^{(\alpha)} \right) = 0
\label{eq:residual_app}
\end{equation}

The system of equations for all slip systems ($N_{\text{sys}}$) is solved using a local Newton-Raphson method. The iterative update for the slip rates is:

\begin{equation}
\sum_{\beta=1}^{N_{\text{sys}}} \mathcal{J}_{\alpha\beta} \delta \dot{\gamma}^{(\beta)} = -f^{(\alpha)}
\end{equation}

where the local Jacobian $\mathcal{J}_{\alpha\beta} = \partial f^{(\alpha)} / \partial \dot{\gamma}^{(\beta)}$ is expressed as:

\begin{equation}
\mathcal{J}_{\alpha\beta} = \frac{\partial \tau_{t+\Delta t}^{(\alpha)}}{\partial \dot{\gamma}^{(\beta)}} - \frac{\partial \chi_{t+\Delta t}^{(\alpha)}}{\partial \dot{\gamma}^{(\beta)}} - \left[ \left( \frac{\dot{\gamma}^{(\alpha)}}{\dot{\gamma}_0} \right)^m \frac{\partial g_{t+\Delta t}^{(\alpha)}}{\partial \dot{\gamma}^{(\beta)}} + g_{t+\Delta t}^{(\alpha)} \frac{\partial}{\partial \dot{\gamma}^{(\beta)}} \left( \frac{\dot{\gamma}^{(\alpha)}}{\dot{\gamma}_0} \right)^m \right] \text{sgn}(\dot{\gamma}^{(\alpha)})
\end{equation}

where
\begin{equation}
    \frac{\partial \tau_{t+\Delta t}^{(\alpha)}}{\partial \dot{\gamma}^{(\beta)}} \cong - (\mathbf{s}_o^{(\alpha)} \otimes \mathbf{m}_o^{(\alpha)}) : \mathbf{C}_o : (\mathbf{s}_o^{(\beta)} \otimes \mathbf{m}_o^{(\beta)})
\end{equation}
 The $\mathbf{s}_o, \mathbf{m}_o$ are the slip plane normal and slip plane direction in the intermediate configuration. $\mathbf{C}_o$ is the stiffness matrix at intermediate configuration.
\subsection{Hardening and Dislocation Density Derivatives}
The evolution of the Statistically Stored Dislocation (SSD) density governs the isotropic strength $g$. The derivative of the SSD density $\rho_{SSD}$ with respect to the slip rate is given by:

\begin{equation}
\frac{\partial \rho_{SSD}^\alpha}{\partial \dot{\gamma}^\beta} = \delta_{\alpha\beta} \frac{\Delta t \left( \frac{k_M}{b} \sqrt{\rho_{SSD}^\alpha} - k_D \rho_{SSD}^\alpha \right) \text{sgn}(\dot{\gamma}^\alpha)}{1 - \frac{k_M |\dot{\gamma}^\alpha| \Delta t}{2b\sqrt{\rho_{SSD}^\alpha}} + k_D |\dot{\gamma}^\alpha| \Delta t}
\end{equation}

The isotropic hardening derivative is then approximated as:

\begin{equation}
\frac{\partial g_{t+\Delta t}^{(\alpha)}}{\partial \dot{\gamma}^{(\beta)}} = \frac{1}{2} G b \frac{\sum_{\xi=1}^{N} h^{\alpha\xi} \frac{\partial \rho_{SSD}^\xi}{\partial \dot{\gamma}^\beta}}{\sqrt{\sum_{\xi=1}^{N} h^{\alpha\xi} \rho_{SSD}^\xi}}
\end{equation}

\subsection{Kinematic Hardening (Back-stress) Derivatives}
The back-stress $\chi^\alpha$ evolves based on a hardening-recovery law. The derivative with respect to the slip rate used in the Newton-Raphson algorithm is:

\begin{equation}
\frac{\partial \chi^\alpha}{\partial \dot{\gamma}^\beta} \approx \frac{\left[C_1 \delta^{\alpha\beta} - C_2 \chi^\alpha \delta^{\alpha \beta} \text{sgn}(\dot{\gamma}^\alpha) + \chi^\alpha |\dot{\gamma}^\alpha| \frac{\partial C_2}{\partial \dot{\gamma}^\beta} + \chi^\alpha \frac{\partial C_3}{\partial \dot{\gamma}^\beta}\right] \Delta t}{\delta^{\alpha \beta} + (C_2 |\dot{\gamma}^\alpha| - C_3) \Delta t}
\end{equation}

\section{Global Tangent Stiffness Matrix}
\label{app:tangent_stiffness}

Finite element solvers such as Abaqus (UMAT) or MOOSE require the algorithmic tangent stiffness tensor $\mathbb{C}^{alg}$, defined as:

\begin{equation}
\mathbb{C}^{alg} = \frac{\partial \Delta \boldsymbol{\sigma}}{\partial \Delta \boldsymbol{\varepsilon}} \approx \frac{\partial \dot{\boldsymbol{\sigma}}}{\partial \mathbf{D}}
\end{equation}

By differentiating the hypoelastic constitutive relation $\dot{\boldsymbol{\sigma}} = \mathbf{C} : (\mathbf{D} - \mathbf{D}^{P})$, we obtain:
where $\mathbf{C}$ is the fourth rank elastic stiffness tensor in the current configuration.
\begin{equation}
\mathbb{C}^{alg} = \left[ \mathbf{C}^{-1} + \Delta t \left( \frac{\partial \mathbf{D}^P}{\partial \boldsymbol{\sigma}} \right) \right]^{-1}
\end{equation}

The sensitivity of the plastic deformation rate to the stress $\partial \mathbf{D}^P / \partial \boldsymbol{\sigma}$ is evaluated utilizing the inverse of the local Jacobian $\mathcal{J}_{\alpha\beta}$ and the Schmidt projection tensors $\mathbb{P}^{(\alpha)}$:

\begin{equation}
\frac{\partial \mathbf{D}^P}{\partial \boldsymbol{\sigma}} = \sum_{\alpha = 1}^{N_{\text{sys}}} \sum_{\beta = 1}^{N_{\text{sys}}} \mathbb{P}^{(\alpha)} \left[ \mathcal{J}_{\alpha\beta} \right]^{-1} \mathbb{P}^{(\beta)}
\end{equation}

This implicit formulation ensures quadratic convergence of the global equilibrium iterations and provides numerical stability for complex thermomechanical fatigue simulations.


\bibliographystyle{elsarticle-num}
\bibliography{references}
\end{document}